\documentclass[conference,compsoc]{IEEEtran}

\usepackage[T1]{fontenc}
\usepackage[utf8]{inputenc}
\usepackage{hyphenat}
\usepackage{xspace}
\usepackage{comment}
\usepackage{xcolor} 
\usepackage{pgf}

\usepackage{booktabs}
\usepackage{multirow}
\usepackage{tabularx}
\usepackage{ragged2e}
\usepackage{supertabular}
\usepackage{makecell}
\usepackage{adjustbox}
\usepackage{changepage}

\usepackage{caption}
\usepackage{subcaption}

\usepackage{float}  
\usepackage{capt-of} 

\usepackage{siunitx}
\usepackage{paralist}

\usepackage{fancyvrb}
\usepackage{framed}
\usepackage{resources/snippets/pygments/pygmentize}

\usepackage[normalem]{ulem}
\usepackage{balance}

\usepackage[square,sort,comma,numbers]{natbib}

\usepackage{xurl}
\usepackage{hyperref}
\hypersetup{colorlinks={true},linkcolor={blue},citecolor=green,urlcolor=blue}

\usepackage{cleveref}

\usepackage{listings}

\lstdefinelanguage{JavaScript}{
  keywords={break, case, catch, continue, debugger, default, delete, do, else,
    finally, for, function, if, in, instanceof, new, return, switch, this,
    throw, try, typeof, var, void, while, with, let, const},
  keywordstyle=\bfseries,
  sensitive=true,
  comment=[l]{//},
  morecomment=[s]{/*}{*/},
  morestring=[b]',
  morestring=[b]"
}

\newcommand{\ie}{i.e.,}
\newcommand{\eg}{e.g.,}

\newcommand{\para}[1]{\smallskip \noindent \textbf{#1}}

\newfloat{listing}{tbp}{lol}
\floatname{listing}{Listing}

\newif\ifdraft

\draftfalse

\ifdraft
\newcommand{\sandra}[1]{{\color{purple}Sandra: #1}}
\newcommand{\sa}[1]{{\color{teal}#1}}
\newcommand{\sas}[1]{{\color{red}Shubham: #1}}
\newcommand{\salim}[1]{{\color{blue}Salim: #1}}
\newcommand{\empirical}[1]{\setlength{\fboxsep}{1pt}\fbox{#1}}
\newcommand{\numemp}[1]{\empirical{\num{#1}}}

\else
\newcommand{\sandra}[1]{}
\newcommand{\sa}[1]{#1}
\newcommand{\sas}[1]{}
\newcommand{\salim}[1]{}
\newcommand{\empirical}[1]{#1}
\newcommand{\numemp}[1]{\num{#1}}
\fi

\newcommand{\pg}{\textit{PageGraph}\xspace}
\newcommand{\Countries}{5\xspace}
\newcommand{\CountryNames}{\textit{Algeria, Germany, India, the UAE, and the USA}\xspace}

\newcommand{\localstorage}{local storage\xspace}
\newcommand{\vpn}{NordVPN, Windscribe, and VyprVPN\xspace}

\begin{document}

\title{\textit{``Where You Are Changes What You Get'':}~\\How Geography Shapes Web Security and Privacy}


\author{
\IEEEauthorblockN{Salim Chouaki}
\IEEEauthorblockA{\textit{New York University Abu Dhabi}}
\and
\IEEEauthorblockN{Shubham Agarwal}
\IEEEauthorblockA{\textit{Max Planck Institute for Security and Privacy}}
\and
\IEEEauthorblockN{Sandra Siby}
\IEEEauthorblockA{\textit{New York University Abu Dhabi}}
\and
}

\maketitle

\begin{abstract}
Web applications underpin essential global and regional services, yet their security and privacy (S\&P) guarantees may differ across regions. That is, users browsing the same website from different countries may experience different risks.
Unfortunately, our understanding of the Web across geographies is skewed, as prior studies largely focused on a) globally ranked websites alone and b) a single vantage point. They may therefore misrepresent the S\&P guarantees and the risks that Web users actually experience.

To address this, we investigate how geography influences the client-side S\&P posture of Web applications. First, we propose a sampling strategy that includes both global and regionally popular websites for individual countries. Then, we observe different client-side S\&P signals from these websites -- deployed HTTP headers, 3\textsuperscript{rd}-party JavaScript inclusion and execution, 3\textsuperscript{rd}-party tracking, and browser fingerprinting -- through visiting them on physically located machines.
Our results reveal significant variation in users' S\&P experiences across geographies. 
The same set of global websites treats users in different countries differently: nearly 1/4 third-party script inclusions on global websites differ depending on the user's country, only 38–50\% of websites load the same 3\textsuperscript{rd}-party trackers for any two countries, and WebRTC fingerprinting increases by 15× for some countries over others.
Likewise, regional websites deploy security-related HTTP headers at rates comparable to global websites, but their effective policies diverge sharply. They also embed regional third parties that global websites never load.

Our findings reveal substantial geographic influence on Web S\&P measurements, beyond ``server-side lottery'' effects. Finally, we discuss divergent trends, their potential causes, and suggestions for improved future measurements.
\end{abstract}

\section{Introduction}
\label{sec:intro}

\sa{The web has grown increasingly \textit{accessible}, \textit{dynamic}, and \textit{localized} -- attracting users worldwide, adapting to their devices, and personalizing content to their needs. But this personalization has a hidden side. Imagine opening your favorite news portal, say \url{theguardian.com}, on a trip across countries and continents: the headline looks the same, the layout looks the same -- yet behind the scenes, the page may be silently talking to a single tracker in one country and over 50 in another. Websites routinely tailor not just what they show you, but what they load and share about you, based purely on where you happen to be. This geography-based dynamism, while often invisible, raises significant security and privacy (S\&P) concerns for users.


The scientific community has extensively examined the security and privacy threats lurking on the web. Researchers have measured the prevalance of stateful tracking~\cite{acar2014web, englehardt2016online, munir2023cookiegraph} and browser fingerprinting~\cite{acar2014web, englehardt2016online, iqbal2021fingerprinting}. Security studies focused on website vulnerabilities, security header configurations, and the integrity of third-party scripts~\cite{van2014large, calzavara2020tale, so2023more}.
Further work also evaluated tracking protections~\cite{amjad2023blocking}, privacy policies~\cite{liu2002raising, moscato2013privacy, zimmeck2014privee} and consent mechanisms~\cite{degeling2018we, liu2022opted, ogut2024dissecting, jha2022internet} deployed by websites.
Recently, researchers also vested efforts into studying these phenomena in a broader scope through studying cross-country differences in third-party tracking on news websites~\cite{hu2020multi, vekaria2021differential}, understanding filter-list coverage across regions~\cite{bottger2025understanding}, investigating how consent banners function across different countries~\cite{ogut2024dissecting}, and comparing privacy practices of healthcare websites in USA, UK and India~\cite{balde2023comparative}. 
These studies are important steps toward understanding a global view of the web. However, all these top-tier research studies share a common measurement design which has various shortcomings:
(1) They focus only on globally popular websites, or a narrow category of regional websites such as news~\cite{vekaria2021differential} or healthcare~\cite{balde2023comparative}. Websites that serve large local audiences (\eg~\url{usps.com} in the USA, or \url{gmx.net} in Germany) may be absent from these top N lists, and their third-party dependencies and security configurations may differ from those of globally popular websites.
(2) They collect data via VPNs or cloud tunnels (e.g., Amazon S3, Azure). As we later demonstrate, and noted by Böttger et al.~\cite{bottger2025understanding}, such services are frequently server-side fingerprinted by websites, resulting in capturing behavior that deviate markedly from the users' reality.
(3) prior regional measurements focused on a limited set of indicators (\eg~presence of cookie banners or privacy policies) which only paints a  view of how client-side S\&P signals vary across regions -- websites load resources and execute third-party scripts at runtime, and these scripts may condition their behavior on the visitor’s location.

In this study, we address these challenges through conducting a representative investigation of client-side S\&P configurations of global and regional websites, and across different geographic locations. We consider physical vantage points in five countries with distinct regulatory regimes and web ecosystems: \CountryNames. We propose an inclusive website sampling strategy to capture a more realistic behavior of the web across countries. To this end, we build three catalogs of websites: \textit{i.)}~globally popular sites, identical for every country (\emph{(D1)}),
\textit{ii.)}~country-coded versions of global services 
(e.g., amazon.de vs. amazon.com)
(\emph{D2}), and \textit{iii.)}~websites that are specific to individual countries \emph{(D3)}.
This way, \emph{D1} allows us to measure how the same website treats users in different locations, while comparing \emph{D1} with \emph{D2} and \emph{D3} reveals how local websites and deployments differ from the heavily-studied global ones.
We then analyze various client-side S\&P signals
(\eg~third-party script inclusions, user tracking, etc.) 
that these websites exhibit and report variances caused due 
to geographic influence.

Our findings show that both the vantage point and the website catalog shape the S\&P behavior the end user experiences. 
The same globally popular website may treat users in different countries differently. 7.5\% of \emph{D1} websites serve different \textit{Content-Security-Policy} depending on the visitor’s location, and the third-party scripts they include differ in vendor origin and inclusion depth. 
Privacy practices diverge more sharply: for any two countries, only 38–50\% of \emph{D1} websites load the same set of trackers, and 15–18\% load entirely disjoint sets. WebRTC-based fingerprinting on these websites is 11–15× more prevalent from Algeria or the UAE than from Germany or the USA.
Then, local websites also show stark deviation from global trends. They adopt security headers at similar rates but with markedly different effective policies. Indian government websites, for instance, explicitly deny access to camera, microphone, and geolocation, while UAE government websites allow it by default. They also embed regional third parties that global websites never load. 

Together, these findings show that website selection and accessors' geography play a pivotal role in the end S\&P guarantees they enjoy. Global rankings overlook local website behavior, while single-location observations miss geographic variation within the same websites. Our study combines both dimensions and provides recommendations for incorporating them into future web measurements.}

To summarize, our main contributions are as follows:
\sa{\begin{compactitem}

    \item We propose a framework to construct a diverse dataset of websites for S\&P measurements that captures both global and regional web experiences and are representative of the \textit{World Wide web}.

    \item We present a large-scale, multi-vantage-point study of Web applications and their client-side S\&P controls across \Countries~countries, accessed under local conditions.

    \item
    We analyze multiple client-side S\&P signals and show that both website type (global or regional) and the users' location are critical factors that shape the end security and privacy boundaries.
    
    \item Based on these findings, we provide recommendations to the research community and industry on conducting more representative web measurements in the future.
\end{compactitem}
}

\section{Background \& Related Work}
\label{sec:techbg}

We first describe the S\&P aspects of web applications that we measure in this work. We then review related studies that reported regional S\&P observations and practices.


\subsection{Client-side Security}
\label{sec:sec-mech}
\para{Security Headers:}~Modern browsers allow applications to enforce security policies and restrictions on the client side through HTTP response headers. The server sends the policy with the response, and the browser applies it before rendering. More than \numemp{10} such headers exist~\cite{owasp-headers}. For instance, \textit{Content-Security-Policy} allows to block all unwanted third-party JavaScript through selectively allow-listing the required ones, thus mitigating XSS-related vulnerabilities.
\textit{CORS} headers restrict \textit{cross-site} resource access. 
\textit{Permissions-Policy} controls which embedded content may request sensitive permissions (\eg~\textit{camera}). 
\textit{Strict-Transport-Security} enforces encrypted connections. 
Unfortunately, security misconfigurations are common, and developers find it difficult to write correct policies~\cite{roth2020complex, calzavara2020tale}.

\para{JavaScript (\&) Vulnerabilities:}~
Web applications may be susceptible to client-side vulnerabilities such as Cross-site Scripting (XSS), Cross-site Request Forgery (CSRF), and clickjacking, which may jeopardize the integrity of the application. Most of these stem from insecure or untrustworthy third-party JavaScript that the page includes and that executes in the page's trusted context~\cite{stock2015facepalm, lekies2015unexpected}.
A malicious script may steal sensitive cookies (cookie theft), overwrite them (session fixation), or perform other illegitimate actions (CSRF, for instance) on behalf of the user. 
Defenses exist, but applications find them hard to adopt because they depend heavily on third-party code~\cite{steffens2021s, so2023more}, and inconsistent development practices widen the attack surface further~\cite{van2014large, lim2023longitudinal, bhuiyan2025non}.

\sa{While the researchers in above studies measured Web applications at large and reported various security vulnerabilities, lapses and inadequate defenses, none of the prior studies accounted for geography to play a role in influencing their observations. The closest study to ours is the \textit{security lottery effect} reported by Roth et al.~\cite{roth2020complex} which showed that Web servers may send different HTTP headers across distinct visits and client configurations (\eg~user-agents)}.

\subsection{Privacy on the Web}
\label{sec:priv-mech}

\para{Stateful Tracking:}~Stateful tracking refers to tracking techniques that rely on browser storage, such as cookies and \localstorage, to maintain state across page visits and sessions. 
First parties and embedded third parties use it for functionality, analytics, advertising, and cross-site tracking.
Early studies measured stateful tracking at scale by analyzing browser storage and distinguishing tracking values from benign state through their structure, persistence, and entropy~\cite{acar2014web, englehardt2016online, lerner2016internet, randall2022measuring}. 
Later work used machine learning to detect tracking-related first-party cookies~\cite{munir2023cookiegraph}. 
Other work used curated blocklists such as EasyList~\cite{easylist} and EasyPrivacy~\cite{easyprivacy} to measure third-party tracking~\cite{yu2016tracking, libert2015exposing, mayer2012third}.

\para{Browser Fingerprinting:}~Browser fingerprinting is a tracking technique that combines device- and browser-specific attributes, collected through JavaScript APIs, into a high-entropy identifier that recognizes a browser across visits and websites~\cite{eckersley2010unique, boussaha2024fp}. Many APIs contribute to such identifiers. Canvas fingerprinting exploits differences in image rendering and font availability; AudioContext fingerprinting exploits variations in audio processing; and WebRTC fingerprinting exposes local and public IP addresses~\cite{eckersley2010unique, mowery2012pixel, acar2014web, englehardt2016online, iqbal2021fingerprinting}.
Acar et al.~\cite{acar2014web} found canvas fingerprinting on over 5\% of the top 100K websites, and Englehardt et al.~\cite{englehardt2016online} on 1.6\% of the top 1M. Iqbal et al.~\cite{iqbal2021fingerprinting} detected fingerprinting from JavaScript execution traces and reported it on more than 10\% of the top 100K websites.
Unlike stateful tracking, browser fingerprinting does not rely on storing identifiers on the client. It is therefore harder to detect, and may persist despite browser restrictions on cookies and storage~\cite{10.1145/3696410.3714699}.

\sa{As with existing security literature, the above studies establish client-side privacy issues on the Web, however, without considering the location-driven variances that may affect geographically-dispersed users differently. Research on differential deployment and behavior of cookie-consent banners across countries is the closest example to ours.}

\subsection{S\&P Practices across Geographies}
\label{sec:users-dev}
\sa{Contrary to the long-tail of research, recent studies included conducting S\&P investigations from more than one location.} Several studies measured third-party tracking on sites popular in different countries: through AWS tunnels on the top 10K sites of nine different countries~\cite{bottger2025understanding}, from two university networks on 10K sites from 20 different countries~\cite{bhuiyan2025digital}, and from the browsing histories of 16 users in four different countries~\cite{hu2020multi}. Similarly, as discussed before, Roth et al.~\cite{roth2022security} measured inconsistencies in security header deployment on the top-10K sites across VPN exits in 218 countries. Others studies focused on isolated web contents and components, such as consent banners~\cite{ogut2024dissecting}, news media~\cite{vekaria2021differential}, and privacy policies~\cite{balde2023comparative} across countries.

Our study differs in two distinct ways. First, most prior work utilizes cloud or VPN infrastructure~\cite{bottger2025understanding, roth2022security, ogut2024dissecting}, which websites detect and treat substantially different from residential traffic (as we later demonstrate in Section~\ref{sec:vpn-exp}). We visit websites from physical machines on local networks in each country. Second, no prior study holds the same websites fixed across local vantage points while also measuring locally popular websites. We investigate both, which separates how a website treats users in different locations from how local websites differ from global ones.


\section{Research Methodology}
\label{sec:method}

\begin{figure*}
	\centering
	\includegraphics[width=0.75\textwidth]{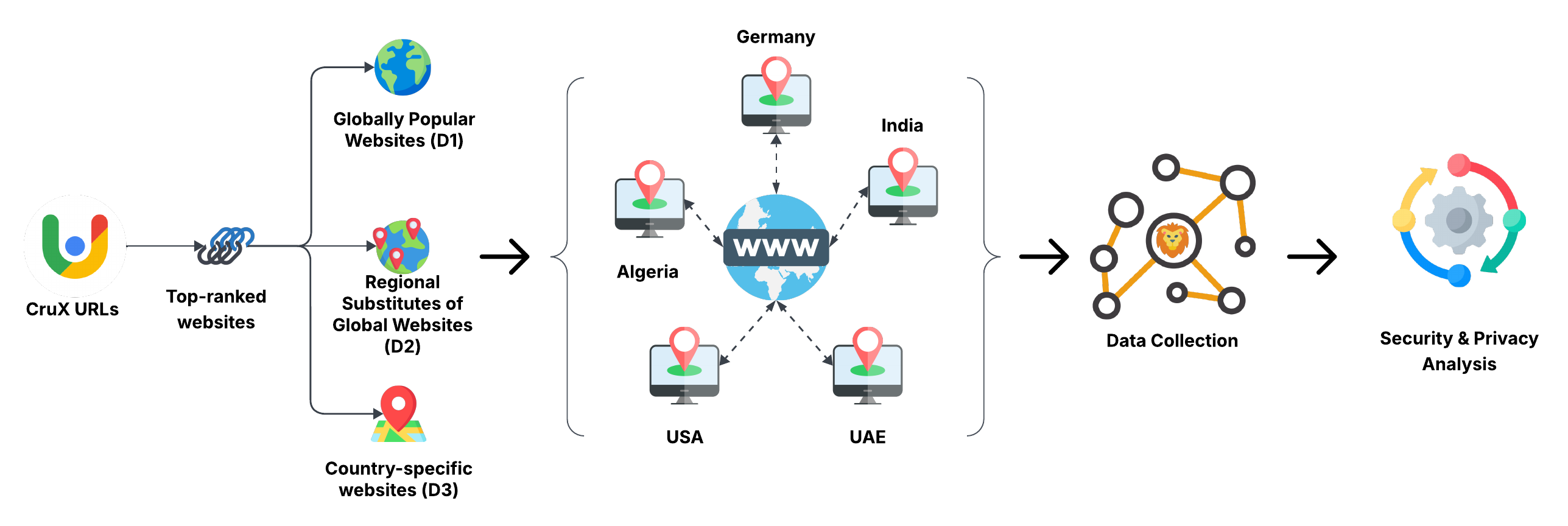}
	\caption{An overview of our URL sampling strategy, automated data collection, and analysis pipeline.}
	\label{fig:method}
\end{figure*}

In this section, we describe our research methodology and experimental setup to measure the S\&P \textit{behavior} of websites across geographies. We study websites that are popular and accessible from \Countries countries: \CountryNames (\Cref{sec:vantage-point-config} gives the rationale for selecting these countries). \Cref{fig:method} illustrates our pipeline. 
We first detail our website cataloging strategy, which provides a fair representation of global and local websites. We then describe our data collection infrastructure and the S\&P signals we measure.


\subsection{Website Categorization \& Cataloging}
\label{sec:url-dataset}

We consider three distinct catalogs of websites per country: (1) globally popular websites, (2) country-coded versions of globally popular websites (\eg~\path{amazon.in} for India), and (2) websites that are primarily popular within the country. We describe the cataloging strategy for our \Countries countries, however, the strategy applies to any country.


\para{General Strategy.}~We construct these catalogs based on the top-level domains (TLDs) of websites. We build the list of TLDs associated with each country from two sources. First, we extract from the IANA Root Zone Database~\cite{iana_rootdb} the TLDs listed as country-code: \texttt{.dz} for Algeria, \texttt{.de} for Germany, \texttt{.in} for India, \texttt{.ae} for the United Arab Emirates, and \texttt{.us} for the United States. 
However, this database does not cover recently proposed geographic generic TLDs and therefore misses region- and city-specific TLDs (e.g., \texttt{.abudhabi} is listed as a generic TLD). To complement it, we use the ICANN gTLDs~\cite{icannwiki_gtld}, which record applications for new geographic top-level domains. \Cref{tab:etld-labels} lists the resulting eTLDs for each country.


\para{Globally-popular websites (\emph{D1}).}
This catalog consists of globally popular websites. We visit the same \emph{D1} sites from every vantage point to investigate whether users in different regions face S\&P risks on the same websites.

We fetch the top \numemp{10000} global sites from CrUX~\cite{crux}. Some websites appear in this list under multiple TLDs (\eg~\path{amazon.com}, \path{amazon.in}, and \path{amazon.de} are all in the global top \numemp{1000}). We exclude the versions whose TLD belongs to one of our countries (\eg~\texttt{.de}, \texttt{.in}) and keep only the global version (\path{amazon.com}). We apply this filter because country-coded counterparts of global sites serve explicitly to users of one country, and rank highly due to local Internet traffic, not global popularity: \path{amazon.in} is in the global top \numemp{1000} primarily due to its traffic from India~\cite{semrush_amazon_in_overview}. We analyze these versions separately in \emph{D2}. After applying this filter, \numemp{7761} sites remain from the top \numemp{10000}. We complement them with \numemp{2239} websites randomly sampled from the global top \numemp{50000}. applying the same exclusion criteria.

\para{Country-coded websites (\emph{D2}).} 
This catalog contains country-coded versions of global websites. We include these to assess
whether users of country-coded websites (e.g., \path{amazon.de}), however, managed by same global parent (e.g., \path{amazon.com}) encounter different S\&P risks.

We first build a candidate set of country-coded versions of global websites. We fetch the top 1 million sites for each country, and the top 1 million global sites from CrUX. 
A site becomes a candidate for a country if (1) it is not in \emph{D1}, (2) it uses one of the country’s eTLDs, and (3) its second-level domain appears in the global list under a different eTLD. The third condition ties the candidate to a global service: \path{example.in} qualifies for India only if its global counterpart, such as \path{example.org} appears in \emph{D1}.

We then validate candidates to eliminate false matches and ensure that each site in the \emph{D2} catalog is likely a country-coded version of a global service (e.g., \path{aadl.dz} and \path{aadl.com} share a second-level domain but belong to different entities). 
We compare two signals between each candidate and its global site: \emph{favicons}~\cite{HTML_Favicon} and \emph{TLS certificates}~\cite{TLS_Certificates_RFC5280}. Favicons provide a visual identifier of a website's brand; similar favicons suggest a shared brand or service. 
We collect the favicon of each candidate and its global site, and compute their visual similarity using embedding-based image comparison. 
For TLS certificates, we compare Subject Alternative Names (SANs) and organization information. SANs list the domain names a certificate covers and therefore indicate whether the candidate and its global site share a certificate. Organization fields identify the entity a certificate was issued to, and therefore allude towards shared ownership. We collect TLS certificates for all candidates and their global sites and compare them. 
We retain a candidate if either signal validates it. 
Appendix~\ref{secc:app-d2-construction} provides more details.


\para{Country-specific websites (\emph{D3}).}  
This catalog includes websites that serve users of one country but are not popular globally. These sites are most likely developed and maintained within the country. They therefore reveal the S\&P practices and development patterns of of local web ecosystem. 

For each country, we fetch its top 1 million websites from CrUX and exclude any site (hostname) already included in \emph{D1} or \emph{D2}. This ensures that websites in \emph{D3} are predominantly popular within the country but not globally.
We then retain only websites that use one of the country's eTLDs. Some local websites may use generic eTLDs, but we conservatively exclude them to keep a clear distinction between globally popular sites and those intended primarily for local audiences. Finally, we select the top-ranked \numemp{10000} from the remaining sites to form the \emph{D3} catalog.

Table~\ref{tab:sites_per_bucket} presents the website counts across catalogs and countries. \sa{Notably, since our aim in this study is to characterize geographic influence on the S\&P behavior of web applications. Hence, we apply a best-effort strategy to build sufficiently comparable \emph{D2} and \emph{D3} catalogs for all countries instead of aiming for comprehensive analysis of the entire web.} Their sizes still differ across countries, because countries differ in how widely their TLDs are used. A small \emph{D2} means that few global services operate a country-coded domain for that country. A small \emph{D3} means that few local sites use the country's TLDs. We resorted to this conservative strategy to avoid false positives; the approach can be relaxed under a tolerable false-positive rate.

\begin{table}[!t]
\centering
\scriptsize
\caption{Number of websites per catalog and per country}
\label{tab:sites_per_bucket}
\begin{tabular}{l|c|c|c|c|c}
\toprule
\textbf{Catalog} & \textbf{Algeria} & \textbf{Germany} & \textbf{India} & \textbf{UAE} & \textbf{USA} \\
\midrule
D1 & \multicolumn{5}{c}{\numemp{10000}} \\ \midrule
D2 & \numemp{65} & \numemp{4278} & \numemp{1535} & \numemp{505} & \numemp{371} \\ \midrule
D3 & \numemp{2981} & \numemp{10000} & \numemp{10000} & \numemp{9749} & \numemp{9122} \\ 
\bottomrule
\end{tabular}
\end{table}


\subsection{Data Collection Overview}
\label{sec:data-collection}

We crawl all \emph{D1} sites from every country, and each country's \emph{D2} and \emph{D3} sites from that country. We visit them through \pg~\cite{pagegraph}, a web measurement tool integrated into the Brave browser~\cite{brave_browser}. \pg records page execution as a directed graph in which nodes are runtime entities and events (scripts, DOM elements, JavaScript calls, storage operations), and edges are the relationships between them (element creation, script inclusion, network requests). This representation allows to attribute every recorded action to specific scripts and execution contexts (\eg~which script performed a given request). \pg does not record HTTP headers~\cite{pagegraph}. Hence, we also record HTTP Archive (HAR) files. HAR files include request and response headers and all other metadata associated with network requests.



We disable \textit{Brave Shields} for all crawls (\ie~all additional privacy protections turned off) to capture organic web activity~\cite{braveshields}. Since Brave is Chromium-based, this setup provides an environment similar to that of most Internet users~\cite{statcounter_browser_market_share}.
For each site, we attempt to visit up to three pages to capture behavior on the landing page and on further pages it embeds~\cite{10.1145/3419394.3423626}. We start with the landing page and iteratively select random same-site URLs; we stop early if we find no such links.
Following prior work, we apply a best-effort strategy to accept cookie consent banners on each page~\cite{moti2024targeted}.
Our automated visits are stateless across websites: we create an individual browser profile for each website (including sub-pages) to avoid bias from state carried over across different websites.
To mitigate any server-side randomness, we visit each URL three consecutive times with distinct browser profiles and report a behavior only if we observe it in all three visits.
Lastly, we disable client-side caching via Puppeteer to ensure that all resources are fetched afresh for each page visit~\cite{puppeteer_setCacheEnabled}.


\subsection{Privacy and Security Measures}
\label{sec:measures}

We utilize the crawl artifacts to analyze websites' S\&P posture across regions. We consider common client-side security configurations (security-related HTTP headers and script inclusion practices) and privacy-invasive behaviors (third-party tracking and browser fingerprinting) to infer geographic influence here. For the former, our goal is not to identify specific security vulnerabilities among websites. Instead, following prior work, we focus on a set of security indicators that are not tied to specific vulnerabilities, but rather provide an overview of differences in the development and deployment of web applications across locations. \sa{As with website catalogs, we do not aim to examine \textit{all} possible client-side S\&P configurations, but rather focus on most prevalent set of signals here for our study.}


\para{Client-side Security Constraints.}
Web applications may restrict their own client-side behavior through security policies that servers send as HTTP response headers. Headers such as \textit{Content-Security-Policy} and \textit{Strict-Transport-Security} play a pivotal role in mitigating client-side vulnerabilities, such as cross-site scripting, clickjacking, and connection-downgrade attacks. We consider the deployment of various security-related headers, along with their policy configurations, as key indicators of a website's security controls. Notably, the locations we study have no reported government or censorship policies that alter HTTP headers; the headers we observe are set by the applications themselves.
We consider \numemp{22} headers in our data that are recognized to add security or privacy boundaries on the client~\cite{owasp-headers, mozilla-headers}, listed in \Cref{tab:sec-headers}. Following prior work~\cite{mendoza2018uncovering, roth2022security, stock2017web}, we analyze their prevalence across websites and the effectiveness of their policies. We then compare their adoption across regions.

\para{Script Inclusion \& Execution Patterns.}
HTTP headers indicate the security configuration desired by a website. The resources it loads may still undermine it. A website may include scripts from untrustworthy or vulnerable third parties, which compromise its integrity on the client side. For instance, a CSP that trusts a \path{script-src} origin may allow a vulnerable or (later-)compromised script from that origin, which can exfiltrate sensitive data or act on the user's behalf~\cite{weichselbaum2016csp, roth2020assessing}. 
We therefore treat the JavaScript sources a website includes, their runtime behavior, and the CSP directives that govern them as a second signal of its security posture and development practices. 
We capture the URLs of all scripts the visited page embeds, directly or through any same- or cross-origin \textit{iframes}, and all scripts injected dynamically at runtime, through \pg{}. We aggregate these sources per visited site and analyze their inclusions and runtime behavior across geographies. We also record JavaScript API invocations at runtime along with their source (\ie~first- or third-party scripts).


\para{Third-party Tracking.}
Third-party tracking lets services embedded in a page observe users across visits and websites by setting identifiers in the browser and reporting page visits to their own servers.
We identify tracking requests using community-maintained filter lists. We use EasyList~\cite{easylist}, which targets advertising-related domains and resources, and EasyPrivacy~\cite{easyprivacy}, which targets non-advertising trackers such as analytics and user profiling. In addition, we include regional \textit{Affiliated Filter Lists}~\cite{easylist_affiliated_filters}, which capture tracking and advertising resources specific to a language or region: \emph{IndianList} (Indian-language websites), \emph{Liste AR} (Arabic-language websites), and \emph{EasyList Germany} (German-language websites). We merge all lists into one and apply it to all crawls.

\para{Browser Fingerprinting.}
Browser fingerprinting combines device- and browser-specific attributes exposed through JavaScript APIs into a stable identifier. It needs no client-side state and thus persists where cookies are blocked or cleared. We detect browser fingerprinting from JavaScript execution traces. We identify scripts that access browser APIs known to expose device- or browser-specific attributes. 
Our detection logic is rule-based and derived from prior work~\cite{iqbal2021fingerprinting}. We consider several fingerprinting techniques: canvas image, WebRTC, and AudioContext~\cite{iqbal2021fingerprinting,acar2014web}. 
We also match executed scripts, by URL and hash, against the list of known fingerprinting scripts from Iqbal et al.~\cite{iqbal2021fingerprinting}.

\section{Country Selection \& Configuration}
\label{sec:vantage-point-config}

A crucial element of our study is to characterize websites' behavior as close to users' reality as possible. This constraint shapes our country selection and scale. We first explain our rationale on country selection and associated limitations. We then report a small-scale experiment that confirms the influence of VPNs on websites' behavior.

\subsection{Choice of Vantage Points}
\label{sec:country-select}
We selected \Countries countries -- \CountryNames, for two reasons.
First, we select countries that prior work has studied extensively (Germany~\cite{bottger2025understanding, ogut2024dissecting} and the USA~\cite{acar2014web, englehardt2016online}) and understudied countries that carry significant Web traffic (Algeria, India, and the UAE).
Together, the five account for 30\% of global web traffic~\cite{cf-radar, apnic, statcounter}. \sa{Nevertheless, we stress that the S\&P of less globally popular Web applications in low-traffic countries are equally important to their users, hence, our study is extensible to any set of websites and countries.}
Second, our design requires monitoring under ``real-world conditions''. Prior work showed that websites fingerprint VPN traffic and \textit{alter} their behavior in response~\cite{jueckstock2019blind, bottger2025understanding}. We therefore use machines physically located in each country, on local networks. This ensures that websites behave as they would for real users, but it limits our study to countries where we could access such machines. We consider this an acceptable trade-off between representativeness and scale.



\subsection{Influence of VPN Traffic}
\label{sec:vpn-exp}

Motivated by our underlying research question, we conduct a small-scale experiment to investigate \textit{behavioral} differences among websites when accessed over VPN, compared to local network conditions.


We randomly selected \numemp{1500} \emph{D1} sites and visited them from a physical machine in each country and through a VPN with exit node within that country, following the methodology presented in Section~\ref{sec:data-collection}.
We used one vantage point (Germany) to simulate connections from Algeria, Germany, India, the UAE, and the USA using VPNs. 
To ensure that each exit node is physically located in the target country rather than mimicking a virtual address, we use different providers where needed: NordVPN~\cite{nordvpn} for Germany, the USA, and the UAE; VyprVPN~\cite{vyprvpn} for Algeria, and Windscribe~\cite{windscribe} for India.
This way, our setup is not prone to conditional redirection based on the \textit{location} header or partial views due to VPN-driven routing. We disable all optional VPN protections (\eg~tracking and ad-blocking), and keep the default configuration for a uniform setup. We compare four signals between physical and VPN access: (1) the number of network requests, (2) the number of third-party scripts, (3) JavaScript API invocations, and (4) security-related response headers.

Overall, \numemp{713}/\numemp{1157} (61\%) sites accessible from all five countries showed significant differences over VPN in at least one country. The effect of VPNs on the number of network requests differ by country. Websites accessed in the USA and Germany showed negligible effect of VPN (median of 214 and 383 requests/site through physical machine vs. 217 and 372 over VPN, respectively, in the USA and Germany). Sites in India loaded considerably more content over VPN (120 vs.\ 375), which suggests that the VPN bypasses local blocking or filtering mechanisms in place, whereas, the websites in the UAE loaded less content over VPN (429 vs.\ 398). \sa{In either case, measurements over VPN may lead to erroneous S\&P observations on target websites.}

\begin{figure}[t]
    \centering
    \caption{Distribution (gain/loss\%) of categorized JavaScript API invocations on websites when accessed over VPN.}
    \includegraphics[scale=1.0]{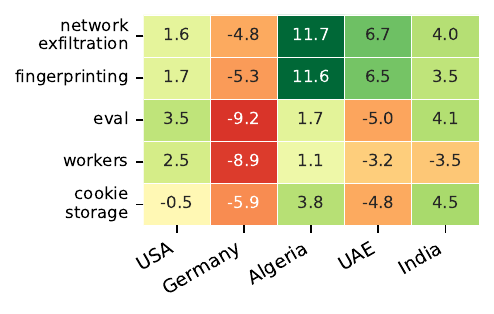}
    \label{fig:js-vpn}
\end{figure}

Third-party scripts follow similar trends (\Cref{fig:script-vpn}). In India, \numemp{72}\% of websites loaded more third-party resources over VPN. For instance, \url{tiktok.com} and \url{preferencenail.com} were included only via VPN on \numemp{42} sites. In Germany, by contrast, VPN access suppressed ad-related scripts such as \url{criteo.net} and \url{adtrafficquality.google} on over \numemp{81}\% of websites. In the UAE, \numemp{69}\% of websites loaded fewer ad-related scripts over VPN, yet recorded more JavaScript invocations (+\numemp{6.5}\%) overall. Some sites even added bot-detection scripts (\eg~\url{cloudflareinsights.com}), The USA was least affected by VPN: some popular ad-related scripts were suppressed while others appeared (\eg~\url{tapioni.com}). 
JavaScript API invocations by these third-party scripts differ accordingly (\Cref{fig:js-vpn}). Sites in India and Algeria showed more tracking (+\numemp{4.5}\%) and fingerprinting (+\numemp{11.6}\%) calls over VPN, in contrast to fewer cases in Germany (-\numemp{6.9}\%).

Security-related headers, by contrast, barely deviated between VPN and local network conditions (\Cref{tab:vpn_headers}). 
Fewer cookies were set via \texttt{Set-Cookie} over VPN, except for a negligible rise in India. We also observed a weak trend of fewer security headers over VPN in the UAE, the USA, and Germany. The VPN provider we used in these three countries could contribute to the observed deviations. Although this analysis is limited to \numemp{1500} globally popular websites, these results suggest that VPN-based measurements may suffice for studying header deployments in isolation.


We do not aim to extensively analyze differences between VPN-based and non-VPN measurements, the reasons behind these discrepancies, or the performance of specific VPN services. Nevertheless, these results show that VPN-based measurements offer a limited perspective compared to users' experience. This motivated our decision to rely on physical machines in each location.


\section{Evaluation \& Results}
\label{sec:results}
We now present the empirical findings from our study across website catalogs accessed from different vantage points.
We investigate variation in enforced security boundaries (security-related HTTP headers and third-party script inclusion) and privacy-invasive behaviors (third-party tracking and browser fingerprinting), as detailed in \Cref{sec:measures}.

Across all catalogs, we visited \numemp{13046} sites from Algeria, \numemp{24278} from Germany, \numemp{21535} from India, \numemp{20254} from the UAE, and \numemp{19493} from the USA, with a success rate of 86\% overall (81\%--89\% per country). Failures were mainly HTTP 404 responses and timeouts.
Recall that we crawl each site three times. To separate real behavior from any transient server-side or network variations (e.g., non-deterministic server responses~\cite{roth2022security}), we report a geographic difference on a site \textit{only} if we observe it to be consistent across visits.



\subsection{State of Security-related HTTP Headers}
\label{sec:header_results}

\subsubsection{Overall Trends in Deployment}
\label{sec:header_trends}

We consider the presence of security-related HTTP headers (or lack thereof) and their configuration as an indicator of the desired security posture of individual websites. Hence, we identify the most common security-related response headers and report their adoption across countries and catalogs. Importantly, we consider a header deployed by a website only if it consistently appears across all three visits to the corresponding sites. \Cref{tab:d1_headers} presents the distribution of the \numemp{10} headers.


\begin{table*}[!ht]
\centering
\scriptsize
\caption{Distributions of Top 10 security-related header deployments across different countries and catalog of websites (D1, D2 and D3). Please note that the distribution is computed where the denominator is the number of eTLDs for each country where at least one request was successful (\eg~response status code was 200)}.
\label{tab:d1_headers}
    \begin{tabular}{l|c|c|c|c|c|c|c|c}
    \toprule
        HTTP Headers & URL Categories & ALGERIA & GERMANY & INDIA & UAE & USA & Mean & Std. Dev \\ \midrule
        \multirow{3}{*}{Strict-Transport-Security} & D1 & 38.93\% & 40.71\% & 34.47\% & 41.14\% & 24.02\% & 35.85\% & 7.12\% \\
        ~ & D2 & 41.82\% & 47.25\% & 31.35\% & 53.58\% & 28.72\% & 40.54\% & 10.50\% \\
        ~ & D3 & 19.79\% & 45.63\% & 36.00\% & 32.51\% & 37.58\% & 34.30\% & 9.43\% \\ \hline
        \multirow{3}{*}{X-Content-Type-Options} & D1 & 31.83\% & 33.27\% & 29.04\% & 33.89\% & 19.33\% & 29.47\% & 5.97\% \\
        ~ & D2 & 41.82\% & 44.59\% & 31.44\% & 47.56\% & 19.68\% & 37.02\% & 11.44\% \\
        ~ & D3 & 29.47\% & 47.45\% & 39.18\% & 29.87\% & 39.28\% & 37.05\% & 7.53\% \\ \hline
        \multirow{3}{*}{X-Frame-Options} & D1 & 32.21\% & 34.19\% & 28.76\% & 34.40\% & 19.64\% & 29.84\% & 6.13\% \\
        ~ & D2 & 32.73\% & 41.68\% & 27.04\% & 45.56\% & 15.96\% & 32.59\% & 11.81\% \\
        ~ & D3 & 27.13\% & 40.81\% & 40.54\% & 28.10\% & 33.22\% & 33.96\% & 6.55\% \\ \hline
        \multirow{3}{*}{X-XSS-Protection} & D1 & 23.15\% & 24.16\% & 20.96\% & 25.07\% & 14.00\% & 21.47\% & 4.45\% \\
        ~ & D2 & 25.45\% & 29.46\% & 20.35\% & 32.95\% & 12.77\% & 24.20\% & 7.93\% \\
        ~ & D3 & 21.77\% & 27.51\% & 33.11\% & 19.40\% & 20.58\% & 24.48\% & 5.74\% \\ \hline
        \multirow{3}{*}{Content-Security-Policy} & D1 & 20.49\% & 21.71\% & 18.81\% & 22.11\% & 12.69\% & 19.16\% & 3.84\% \\
        ~ & D2 & 16.36\% & 22.56\% & 23.83\% & 29.80\% & 16.49\% & 21.81\% & 5.62\% \\
        ~ & D3 & 8.41\% & 23.14\% & 25.24\% & 23.99\% & 26.76\% & 21.51\% & 7.45\% \\ \hline
        \multirow{3}{*}{Report-To} & D1 & 20.52\% & 21.09\% & 16.79\% & 20.52\% & 13.49\% & 18.48\% & 3.27\% \\
        ~ & D2 & 1.82\% & 16.01\% & 16.32\% & 19.77\% & 7.45\% & 12.27\% & 7.40\% \\
        ~ & D3 & 1.73\% & 7.12\% & 8.41\% & 12.85\% & 10.03\% & 8.03\% & 4.12\% \\ \hline
        \multirow{3}{*}{Access-Control-Allow-Origin} & D1 & 10.31\% & 11.23\% & 9.43\% & 11.58\% & 6.20\% & 9.75\% & 2.15\% \\
        ~ & D2 & 7.27\% & 9.39\% & 8.16\% & 11.46\% & 6.91\% & 8.64\% & 1.84\% \\
        ~ & D3 & 5.00\% & 6.83\% & 11.67\% & 7.08\% & 5.43\% & 7.20\% & 2.65\% \\ \hline
        \multirow{3}{*}{Permissions-Policy} & D1 & 7.21\% & 7.16\% & 6.37\% & 7.79\% & 4.74\% & 6.65\% & 1.18\% \\
        ~ & D2 & 1.82\% & 7.25\% & 7.24\% & 7.45\% & 12.23\% & 7.20\% & 3.69\% \\
        ~ & D3 & 6.63\% & 6.68\% & 13.01\% & 8.01\% & 5.03\% & 7.87\% & 3.06\% \\ \hline
        \multirow{3}{*}{X-Permitted-Cross-Domain-Policies} & D1 & 2.47\% & 2.74\% & 2.24\% & 2.91\% & 1.42\% & 2.36\% & 0.58\% \\
        ~ & D2 & 1.82\% & 6.09\% & 3.67\% & 8.60\% & 4.26\% & 4.89\% & 2.57\% \\
        ~ & D3 & 3.93\% & 4.66\% & 6.40\% & 4.60\% & 5.96\% & 5.11\% & 1.03\% \\ \hline
        \multirow{3}{*}{Access-Control-Allow-Methods} & D1 & 4.15\% & 4.23\% & 3.75\% & 4.37\% & 2.41\% & 3.78\% & 0.80\% \\
        ~ & D2 & 3.64\% & 2.35\% & 3.21\% & 2.87\% & 1.60\% & 2.73\% & 0.79\% \\
        ~ & D3 & 1.68\% & 2.61\% & 6.77\% & 3.13\% & 2.02\% & 3.24\% & 2.05\% \\ \bottomrule
    \end{tabular}
\end{table*}

Notably, we observe several patterns and anti-patterns across catalogs and countries. To begin with, the adoption of popular headers such as \textit{CSP} (except Algeria and the UAE) and \textit{HSTS} increases from \emph{D1} to \emph{D3}, whereas this is inverse for recent additions (\eg~\textit{COOP} and \textit{Report-To}). \sa{Importantly, this contradicts the prevalent knowledge in the community, where Bhuiyan et al.~\cite{bhuiyan2025digital} reported that CSP adoption is very low and heavily limited globally (across \textit{developed} and \textit{developing} countries). Whereas, we observe differential adoption rate even between "developed" and "developing" countries based on the vantage point, and hence, this broad distinction may not hold in practice.
} Then, the fraction of deployments of individual headers also differ selectively across global websites (\textit{D1}), where we observe a general trend of slightly higher occurrences in the \textit{UAE}, followed by \textit{Germany} and \textit{Algeria} (interchangeable for different headers) --
directing that the security controls are not uniform even across same websites and are heavily dependent on the \textit{visitors'} location.


The differences were more pronounced among \textit{D2} and \textit{D3}, where we see inconsistent trends across individual headers and countries.
In general, \textit{X-CTO} (34.5\%), HSTS (36.9\%), and the (now-deprecated) \textit{X-FO} (34.1\%) headers remain among the prevalent ones across all dimensions. 
In contrast, recently introduced ones, such as \textit{COOP} and \textit{CORP}, are widely visible across \textit{D1}, while \textit{Permissions-Policy} and \textit{CORS} headers are more prominent in \textit{D2} and \textit{D3}. 
\textit{India-specific} sites were an interesting outlier here: of the sites that sent the \textit{ACA-O} header, only \numemp{53.9}\% of them included wildcards (compared to other \textit{D3} sites with \numemp{68}--\numemp{77}\%), while a large fraction of them specifically reflected back the requesting origin to the client.

In summary, the distribution of headers shows that focusing solely on global sites (\textit{D1}) may provide early indicators of deployments, but it does not accurately reflect the state of security header adoption in practice.

\subsubsection{Effective Policy Configurations}
\label{sec:sec_lot_check}

While websites may deploy security-related headers, a weak, misconfigured, or erroneous policy may render them useless. As an example, an \texttt{X-FO} header set to (deprecated) \texttt{ALLOWALL} or a \textit{HSTS} header with \textit{max-age} set to \numemp{0} is effectively ignored by the browser.
Thus, we assess the effectiveness of a subset of these headers by analyzing the values and directives they include. 
We leverage the Roth et al.'s ~\cite{roth2022security} strategy for four popularly deployed headers -- \emph{CSP}, \emph{X-FO}, \emph{HSTS} and \emph{Set-Cookie}. While Roth et al. originally used this strategy to study inconsistencies in header deployment and enforcement across variable client-side configurations (\emph{only}) over VPNs, our goal is orthogonal: we utilize the formal constraints modeled by them to evaluate \textit{consistent} header deployments across global and regional websites, and countries.

We first normalize individual header values before applying the quality checks. When a single response contains multiple \textit{HSTS} headers, we consider only the last occurrence; for \textit{CSP}, we concatenate all values. 
As before, we only consider headers consistently served with identical values across all three visits.
~\Cref{tab:sec-lot-all} summarizes the effective quality assessment for the four individual headers.

CSP deployment and policies varied substantially even when the same sites were accessed from different vantage points. \numemp{7.51}\% of \textit{D1} sites \emph{consistently} serve different CSP policies across locations. Among these, \numemp{4.2}\% differ in the set of directives they included, while \numemp{3.88}\% also deviate in the set of allow-listed sources. We also found that \numemp{3.72}\% of sites varied in terms of enforcement (CSP-RO vs CSP). One notable pattern we saw unique to USA-specific catalogs was that they allow-listed native-app-schemes (\textit{`safari-extension:`, 
chrome-extension:, zoommtg, and rtmp:}) on over \numemp{309} zoom sub-domains (\url{*.zoom.us}), unreported by prior works. \sa{These variances indicate that any measurement study, that monitors prevalence of header controls or characterize security posture of Web applications, are directly affected by their location of observation.}


In addition to the four headers above, we also investigated differential policies in the \emph{Permission-Policy} header, which controls access to sensitive client-side features such as geolocation and sensor APIs. 
In general, using Cloudflare Radar site categorization~\cite{CloudflareDomainCategories}, we found that this header is majorly shipped by adult, gambling, piracy and streaming sites across all catalogs and countries (\eg~\url{ara.xhexperience.xyz} in \textit{D1}, \url{memesapk.co.in} in \textit{D2}, and \url{amankanliga365.us} in \textit{D3}). 
Moreover, we found two notable cross-country cases worth mentioning here: (1) India-specific (\textit{D3}) sites aggressively restricts (\numemp{60.8}--\numemp{79.4}\%) camera, geolocation and microphone API permissions through explicit policy. 
This contrasts with other countries, especially the UAE (\numemp{43}--\numemp{53}\%) and Germany (\numemp{38.2}--\numemp{42.3}\%). 
(2) This trend was even more vivid when only considering government sites: most UAE government sites (\texttt{*.gov.ae}) allowed wildcard access to sensitive APIs, whereas Indian government sites ( \texttt{*.gov.in}) explicitly shipped denying policies to users. \sa{Overall, our findings on global websites align with the trends reported by Fernandez-de-Retana et al.~\cite{fernandez2025permissions}, however, explicit focus on geographic influence and regional sites uncovered unique and deviating deployment configurations.}

These results affirm that header deployment varies across site catalogs and vantage points, and they may differ substantially in their effective configurations. In particular, country-coded (\emph{D2}) and locally popular (\emph{D3}) websites expose policy patterns that are not captured by globally popular websites alone, while even same \emph{D1} websites may serve different policies from different vantage points.


\subsection{Third-party Script Inclusions}
\label{sec:third-party-scripts}

Websites often include JavaScript from numerous third-party sources to integrate services and features on the client side.
However, these third-party JavaScript libraries could become malicious, untrustworthy, or even vulnerable, exposing the application to threats. Thus, we consider the prevalence of these JavaScript sources an additional signal of an application's security posture. We analyze the distribution of first-party vs. third-party JavaScript across websites for each vantage point and website catalog. We define \textit{first-party (1P)} scripts as scripts loaded from both \textit{same-origin} and \textit{cross-origin-but-same-site} sources, while \textit{cross-site} sources are naturally labeled as \textit{third-parties (3P)}. We present the top 10 third-party origins and their prevalence across websites in \Cref{tab:3p_scripts_distrib}. Additionally, we investigate the JavaScript inclusion chain and their execution context to identify third-party \textit{iframes} that include JavaScript. As before, we only considered websites that showed consistent script-related behavior across three distinct visits.

\para{\textit{D1}: Global Websites and Trends.}~Over \numemp{92.7}\% of \textit{D1} sites included at least one third-party script, with a median of \numemp{75} 3P-scripts from \numemp{4} unique domains. This held nearly steady at \numemp{92.4}\% for \textit{D2} sites, however, at a median of \numemp{121} scripts from \numemp{4} domains, but dropped to \numemp{77.0}\% for \textit{D3} sites, with a lower median of \numemp{60} scripts from \numemp{3} domains. This already indicates that the prevalence of third-party scripts is non-uniform across global and local websites.
This trend is also visible at the level of individual 3P-script domains. For instance, \url{googlesyndication.com}, one of the top occurring 3P-script domain appears on \numemp{20.54}\% of \textit{D1} sites, compared with only \numemp{7.23}\% of \textit{D2} and \numemp{5.37}\% of \textit{D3} sites. 
Similarly, \url{criteo.net} appears on \numemp{8.78}\% of \textit{D1} sites, versus \numemp{1.89}\% in (\textit{D2}) and \numemp{1.07}\% in (\textit{D3}), strongly indicating that their presence and behavior on the web may be overrepresented when only focused on a global set of websites. 
We also observe vantage-point differences within \emph{D1}. Over \numemp{8.6}\% of global websites load scripts from \numemp{10} or more third-party domains based on the point of access. For example, the scripts loaded from the \textit{AliExpress} sub-domains  (\url{*.aliexpress.com} appeared in Germany and the UAE more frequently (75\%--95\% share) and with high recursive inclusion depth (5) than what we observe in the USA (33\% share) with only direct inclusions. \sa{Hence, observing websites, their supply-chain units and vendors for S\&P or compliance checks may very much depend on the source location of the observer.}

\textit{The UAE Anomaly:}~We expand our analysis of 3P-script request volume per website by taking the vantage point into consideration, as detailed in \Cref{tab:3p_script_volume_distrib}. Here, we see clear distinction in trends based on the point of access, as well as the website catalogs. 
The UAE is an outlier in dual directions: the website behavior flips between the high outlier and the low outlier based on the URL category. For globally popular websites (\textit{D1}), visits from the UAE trigger higher numbers of 3P-scripts (2.52x compared to Germany) from specific set ad-exchange sources. For example, on \url{9db.jp}, a single 3P-script domain, \url{doubleclick.net}, recursively embedded \empirical{six} more scripts downstream through \texttt{document.write}, which we did not observe in other countries. Whereas, this overall behavior flips the other way for websites in \textit{D2} and \textit{D3}. Websites in Germany loaded 2x ad-related scripts than the UAE counterparts. We speculate and attribute this behavior to exceptionally high Internet penetration rate and unique local ad ecosystem in the region~\cite{uae-ken}.

\sa{Our prevalent understanding on script inclusions and behavior stem from large-scale analyses of millions of top-ranked websites~\cite{stock2017web, steffens2021s, stock2015facepalm}. However, the above findings distorts the existing knowledge on these patterns, the associated behavior (\eg~ads) and threats (\eg~supply-chain vulnerabilities), and the means to measure them at scale.}

\begin{table*}[!ht]
    \centering
    \scriptsize
    \caption{Distribution of Top 10 most prevalent \textit{third-party} (cross-site) scripts included by websites, ranked by mean adoption rate across D1, D2, and D3, across countries and URL categories. The script origins are ordered in descending order of their occurrences across the cumulative sum of websites and across catalogs.}
    \label{tab:3p_scripts_distrib}
    \begin{tabular}{l|c|c|c|c|c|c|c|c}
    \toprule
        \textbf{Script Sources} & \textbf{URL Categories} & \textbf{ALGERIA} & \textbf{GERMANY} & \textbf{INDIA} & \textbf{UAE} & \textbf{USA} & \textbf{Mean} & \textbf{Std. Dev} \\ \midrule
        \multirow{3}{*}{www.googletagmanager.com} & D1 & 59.37\% & 56.4\% & 48.87\% & 59.39\% & 32.74\% & 51.35\% & 11.25\% \\
        ~ & D2 & 56.36\% & 63.44\% & 56.4\% & 71\% & 28.28\% & 55.10\% & 16.16\% \\
        ~ & D3 & 25.54\% & 33.95\% & 32.21\% & 62.18\% & 49.13\% & 40.60\% & 14.83\% \\
        \hline
        \multirow{3}{*}{www.google.com} & D1 & 26.48\% & 26.03\% & 18.94\% & 27.93\% & 13.98\% & 22.67\% & 5.98\% \\
        ~ & D2 & 29.09\% & 19.95\% & 19.96\% & 36.04\% & 14.65\% & 23.94\% & 8.53\% \\
        ~ & D3 & 15.7\% & 10.99\% & 13.37\% & 28.11\% & 27.58\% & 19.15\% & 8.11\% \\
        \hline
        \multirow{3}{*}{googleads.g.doubleclick.net} & D1 & 30.5\% & 24.62\% & 22.03\% & 33.31\% & 15.38\% & 25.17\% & 7.08\% \\
        ~ & D2 & 18.18\% & 20.65\% & 24.98\% & 42.01\% & 11.11\% & 23.38\% & 11.56\% \\
        ~ & D3 & 6.12\% & 12.81\% & 9.88\% & 34.46\% & 11.32\% & 14.92\% & 11.21\% \\
        \hline
        \multirow{3}{*}{connect.facebook.net} & D1 & 20.35\% & 20.08\% & 15.43\% & 21.83\% & 10.75\% & 17.69\% & 4.56\% \\
        ~ & D2 & 38.18\% & 27.65\% & 22.56\% & 41.73\% & 11.62\% & 28.35\% & 12.14\% \\
        ~ & D3 & 9.48\% & 11.23\% & 8.54\% & 23.72\% & 10.79\% & 12.75\% & 6.22\% \\
        \hline
        \multirow{3}{*}{www.gstatic.com} & D1 & 10.93\% & 12.75\% & 7.53\% & 13.07\% & 6.32\% & 10.12\% & 3.06\% \\
        ~ & D2 & 18.18\% & 14.11\% & 13.07\% & 24.93\% & 9.09\% & 15.88\% & 6.01\% \\
        ~ & D3 & 19.47\% & 8.33\% & 11.4\% & 32.58\% & 15.37\% & 17.43\% & 9.45\% \\
        \hline
        \multirow{3}{*}{www.google-analytics.com} & D1 & 17.11\% & 17.09\% & 14.1\% & 18.54\% & 10.08\% & 15.38\% & 3.38\% \\
        ~ & D2 & 16.36\% & 12.14\% & 17.73\% & 19.51\% & 7.58\% & 14.66\% & 4.81\% \\
        ~ & D3 & 11.98\% & 7.95\% & 11.1\% & 16.97\% & 18.66\% & 13.33\% & 4.40\% \\
        \hline
        \multirow{3}{*}{maps.googleapis.com} & D1 & 12.37\% & 13.84\% & 8.7\% & 14.83\% & 7.59\% & 11.47\% & 3.18\% \\
        ~ & D2 & 16.36\% & 9.11\% & 15.04\% & 17.34\% & 7.07\% & 12.98\% & 4.60\% \\
        ~ & D3 & 18.86\% & 7.48\% & 14.5\% & 27.24\% & 25.68\% & 18.75\% & 8.14\% \\
        \hline
        \multirow{3}{*}{cdnjs.cloudflare.com} & D1 & 10.73\% & 22.04\% & 17.58\% & 18.42\% & 14.39\% & 16.63\% & 4.28\% \\
        ~ & D2 & 7.27\% & 12.44\% & 13.52\% & 19.24\% & 7.58\% & 12.01\% & 4.92\% \\
        ~ & D3 & 7.34\% & 4.79\% & 11.65\% & 12.8\% & 12.07\% & 9.73\% & 3.49\% \\
        \hline
        \multirow{3}{*}{static.cloudflareinsights.com} & D1 & 23.63\% & 12.01\% & 19.95\% & 23.02\% & 13.51\% & 18.42\% & 5.38\% \\
        ~ & D2 & 0\% & 5.5\% & 17.99\% & 16.26\% & 10.1\% & 9.97\% & 7.47\% \\
        ~ & D3 & 2.34\% & 2.78\% & 8.71\% & 12.15\% & 15\% & 8.20\% & 5.61\% \\
        \hline
        \multirow{3}{*}{pagead2.googlesyndication.com} & D1 & 20.86\% & 22.18\% & 13.42\% & 24.96\% & 11.76\% & 18.64\% & 5.74\% \\
        ~ & D2 & 7.27\% & 4.94\% & 10.47\% & 9.49\% & 4.55\% & 7.34\% & 2.65\% \\
        ~ & D3 & 2.45\% & 7.21\% & 7.71\% & 2.21\% & 2.83\% & 4.48\% & 2.73\% \\
        \bottomrule
    \end{tabular}
\end{table*}

\para{\textit{D2}/\textit{D3}: Regional Influences on Websites.}~We showed that globally popular websites load different third-party scripts and at different rates, based on the access point. Regional and country-coded websites show similar trends where they even go further to include scripts from regional services over global players.
We saw this explicit trend among websites that loaded AI-assistance related scripts. That is, while only \numemp{1.60}\%--\numemp{2.11}\% \textit{D1} websites loaded them, this starkly increased in \textit{D2} (\numemp{2.45}\%--\numemp{7.8}\%). Interestingly, the UAE sites in \textit{D2} and \textit{D3} were the outliers here as well which included these script the most, compared to other countries (\numemp{7.8}\% and \numemp{2.83}\%). \url{tabby.ai}, for example, is an AI-powered buy-now-pay-later checkout assistant that appeared only on 12 distinct UAE-specific storefronts (\eg~\url{michaelkors.ae} and \url{underarmour.ae,}). Similarly, \url{dynamicyield.com}, an AI-personalization tool, showed up (only) on multiple UAE-specific retail sites (\eg~Swarovski and Lacoste). While German sites also showed elevated inclusion of AI scripts in \textit{D2} (\numemp{3.71}\%), the trend is spread over multiple vendors.

Next, we performed a deeper analysis at the JavaScript activity of the embedded 3P-scripts at runtime and report interesting deviations. Here, we categorized closely-related JS APIs, reported by prior works to be security-critical~\sa{\cite{web-almanac-24, web-almanac-25, snyder2016browser, snyder2017most, iqbal2020adgraph, bahrami2021fp}}. We saw clear downhill activity in eval and service-worker related API calls (\url{onesignal.com} being one prominent case) among \textit{D2} sites when compared to \textit{D1} sites, across all countries. Contrastingly, the fingerprinting and network-related activity (\eg~\url{clarity.ms} and \url{doubleclick.net}) were more prevalent. Irrespective of the observed effective fingerprinting rate, as discussed in \Cref{subsec:fingerprinting}, the associated API calls in the UAE was almost 2.2x more than other countries for \textit{D2} sites. 
Since the findings strongly suggest that third-party scripts and their runtime activity are heavily customized to the accessors' location, merely assessing their true behavior and associated threats and vulnerabilities from a single vantage point may be an underestimation.


Parallel to third-party scripts, we also analyzed the prevalence of 3P-\emph{iframes} and saw mirroring trends. That is, \numemp{82.72}\% \textit{D1}, \numemp{86.86}\% \textit{D2}, and \numemp{80.18}\% \textit{D3} sites loaded at least 3P-iframes. However, the unique 3P-domains counts were 4x higher in \textit{D1} across countries, with highest concentration in the UAE. For example, \url{theguardian.com} included \numemp{278} 3P-iframes in the UAE, while none when accessed from Germany. Furthermore, the iframes among \textit{D2}/\textit{D3} websites heavily skew towards CMP and social-media widgets, as opposed to global websites that concentrates largely around ad-exchange infrastructure. Contrastingly, the USA-local websites included iframes from least diversity of third-parties. For example, \url{consentmanager.net}, \url{usercentrics.eu} and \url{paypal.com} showed up in \numemp{100}+ Germany-specific sites, but none of them appeared in their USA-counterpart.

The above findings corroborate the fact that regional sites may serve as a more potent starting point to measure the web supply-chain and regional deployments. More generally, polymorphic behavior of sites and the resources that they embed also necessitates comprehensive analysis by considering these dimensions for a holistic assessment.

\subsection{Third-party tracking}
\label{subsec:tracking}

We next analyze third-party tracking across site catalogs and countries. A tracker is third-party if its domain (eTLD+1) differs from that of the visited website.


\para{Prevalence of third-party tracking.}
Table~\ref{tab:tracking:summary} summarizes tracking across catalogs and countries. We report tracking on a website if we observe at least one third-party tracker on it.
Third-party tracking is more prevalent, and less variable across countries, on globally popular websites and their country-coded versions than on local websites.
In \emph{D1}, 68.3\%--71.4\% of sites load at least one third-party tracker; in \emph{D2}, 67.1\%--76.0\%. 
\emph{D3}, by contrast, shows markedly lower and far more variable prevalence: it spans 30 percentage points across countries, ten times the spread in \emph{D1}. 
This variation likely reflects differences in local tracking and advertising practices. 
Only 35.0\% of Algerian and 36.1\% of Indian \emph{D3} sites track, compared with 50.8\% of German, 55.4\% of USA, and 65.0\% of UAE sites.

\begin{table}[t]
  \centering
  \scriptsize
  \caption{Overview of third-party tracking across website catalog and vantage point. (a) prevalence (\% of sites with at least one tracker); (b) average number of distinct trackers per site; (c) total distinct tracking domains across all sites.}
  \label{tab:tracking:summary}
  \setlength{\tabcolsep}{4pt}
  \begin{tabular}{cccccc}
    \toprule
    & \textbf{Algeria} & \textbf{Germany} & \textbf{India} & \textbf{UAE} & \textbf{USA} \\
    \midrule
    \multicolumn{6}{l}{\textit{(a) Tracking prevalence (\% of sites)}} \\
    \addlinespace[2pt]
    \quad \emph{D1}  & 71.4 & 70.2 & 68.3 & 71.2 & 70.6 \\
    \quad \emph{D2}  & 73.3 & 74.1 & 71.3 & 76.0 & 67.1 \\
    \quad \emph{D3}  & 35.0 & 50.8 & 36.1 & 65.0 & 55.4 \\
    \midrule
    \multicolumn{6}{l}{\textit{(b) Avg.\ distinct trackers per site}} \\
    \addlinespace[2pt]
    \quad \emph{D1}  & 8.5 & 9.5 &  6.9 & 11.4 & 9.4 \\
    \quad \emph{D2}  &  5.0 &  6.9 &  5.0 &  9.0 &  7.3 \\
    \quad \emph{D3}  &  2.4 &  5.8 &  3.7 &  5.0 &  3.8 \\
    \midrule
    \multicolumn{6}{l}{\textit{(c) Total distinct tracking domains}} \\
    \addlinespace[2pt]
    \quad \emph{D1}  & 1,694 & 1,769 & 1,517 & 1,828 & 1,765 \\
    \quad \emph{D2}  &   52 &  890 &  386 &  299 &  212 \\
    \quad \emph{D3}  &  115 & 901 &  328 &  428 &  549 \\
    \bottomrule
  \end{tabular}
\end{table}


\para{Same websites, different locations.}
The \emph{D1} catalog reveals how tracking differs across locations on the same set of websites. 
We compare the number and set of trackers each site loads from each vantage point.
Location markedly affects both the \emph{volume of tracking} and the \emph{parties} involved. 
The UAE has the most distinct trackers per \emph{D1} site (11.4 on average), followed by Germany (9.5), the USA (9.4), Algeria (8.5), and India (6.9). 
The trackers themselves also differ. For any two countries, only 38.2\%--50.1\% of \emph{D1} sites load exactly the same trackers, and 14.8\%--17.5\% load completely disjoint sets.
Tracking prevalence, however, varies by only 3.1 percentage points across vantage points (Table~\ref{tab:tracking:summary}).
Hence, location has a differential impact on tracking. It barely changes \emph{whether} a globally popular site embeds trackers, but strongly changes \emph{how many} and \emph{which}. We identify recurring cases where certain trackers appear only from specific vantage points. For example, \path{spadsync.com} appears on 187 \emph{D1} sites when loaded from the UAE, but from none of the other four countries.

Pinpointing why a tracker appears from one vantage point but not another is challenging. Several hidden factors may influence this behavior: data-protection regulations, deliberate segmentation by the first party, or conditional execution of third-party scripts. 
We manually inspect selected cases and identify one where the logic is visible in client-side JavaScript. 
When visited from Germany, \path{finanzen.net/ratgeber/garantie-zertifikate/} loads the tracking script \path{script.ioam.de/sensor/latest/manager/base/es6/bundle.js}. 
From Algeria, India, the UAE, or the USA, it does not. 
A first-party script on the page, \path{data-fdbbf15b66.finanzen.net/iomm/latest/manager/base/es6/bundle.js}, includes this tracker only for Germany (\texttt{de}) and Austria (\texttt{at}), with a country-specific endpoint for each (Listing~\ref{fig:code-example}).

\begin{listing}[t]
\caption{Country-based tracker inclusion logic. The script includes different trackers for German and Austrian users.}
\begin{lstlisting}[
      language=JavaScript,
      basicstyle=\ttfamily\scriptsize,
      frame=lines,
      breaklines=true,
      numbers=left,
      numberstyle=\tiny,
      xleftmargin=1.5em
  ]
var Ha = oa(["de", "at"]);
cdn: {
  host: ({ country }) =>
    country === "at" ? "script-at.iocnt.net" : "script.ioam.de"
}
\end{lstlisting}
\label{fig:code-example}
\end{listing}

\para{Country-coded versus global websites.}
We test whether tracking differs between the country-coded version of a service and its global version when both are visited from the same location. We compare each \emph{D2} site with its global equivalent in \emph{D1}, accessed from the corresponding vantage point (e.g., \path{amazon.de} vs.\ \path{amazon.com}, both from Germany). 
Tracking prevalence is high and similar for both: 73.3\% vs.\ 71.5\% in Algeria, 74.1\% vs.\ 70.3\% in Germany, 71.3\% vs.\ 68.4\% in India, 76.0\% vs.\ 71.2\% in the UAE, and 67.1\% vs.\ 70.6\% in the USA.
However, versions of the same service do not load the same trackers, and the country-coded version usually loads fewer: the global version loads more trackers for 38.7\%--53.9\% of services.
Only 12.8\%--28.8\% of country-coded sites load exactly the same trackers as their global equivalents, and 4.0\%--7.7\% load completely disjoint sets. 
Measurements on global versions may therefore miss the trackers that users of country-coded versions actually face.
For instance, the Czech ad network \path{seznam.cz} loads on the global version of 48 German services (e.g., \path{www.tescoma.com}) but on none of their country-coded versions (e.g., \path{www.tescoma.de}).

\para{Tracker ecosystems.}
Table~\ref{tab:tracking:common:trackers} lists the top third-party tracking domains across catalogs. 
Google’s advertising and analytics services dominate from every vantage point: 
\path{googletagmanager.com} appears on 36.4\%--52.5\% of sites, \path{google-analytics.com} on 26.4\%--37.1\%, \path{doubleclick.net} on 18.9\%--35.7\%, and \path{google.com} on 18.3\%--34.0\%.
The size of the tracker ecosystem differs by location. In \emph{D1}, the number of distinct tracking domains ranges from 1,517 in India to 1,828 in the UAE. Interestingly, the differences widen in the regional catalogs: Germany has the most distinct tracking domains (890 in \emph{D2} and 901 in \emph{D3}) and Algeria the fewest (52 and 115).
Some tracking domains appear from a single location only. We find 761 such country-exclusive domains for Germany, 222 for the USA, 154 for the UAE, 140 for India, and 38 for Algeria. The most prevalent ones are localized Google domains: \path{google.ae} (19.9\% of UAE sites), \path{google.dz} (15.2\%), \path{google.de} (14.9\%), and \path{google.co.in} (11.8\%). 
The remaining country-exclusive domains are rare. At most two per country appear on more than 1\% of sites (\eg~\path{sparkasse.de} in Germany, 1.0\%). 
Together, however, they form a long tail of regional trackers. 
In Germany, this tail includes domestic advertising and analytics vendors (\eg~\path{ioam.de}, \path{adcell.com}). In the UAE, it includes Russian services (\eg~\path{bitrix24.ru}, \path{2gis.ru}), and in India, Vietnamese ad networks (\eg~\path{accesstrade.vn}).
Local tracker ecosystems thus reflect regional user bases and ad markets, which single-vantage-point measurements cannot observe.

\subsection{Browser Fingerprinting}
\label{subsec:fingerprinting}

We finally report on the prevalence and geographic variation in \emph{browser fingerprinting} behavior.

\para{Prevalence of browser fingerprinting.}
Table~\ref{tab:fp:all} reports fingerprinting prevalence across website catalogs and vantage points.
As with tracking, fingerprinting is most prevalent in \emph{D1} (9.9\%--16.7\%), lower in \emph{D2} (5.7\%--16.2\%), and lowest in \emph{D3} (0.3\%--5.1\%).
Fingerprinting also differs by location. In \emph{D2} and \emph{D3}, it is highest in the UAE, followed by the USA and Germany, and lowest in India and Algeria. In \emph{D1}, the UAE still leads and India trails, but Algeria ranks second.

\para{Fingerprinting techniques.}
Table~\ref{tab:fp:all} breaks down fingerprinting by technique. Canvas image and audio APIs drive most fingerprinting across catalogs and vantage points. 
Both follow the overall pattern: highest in \emph{D1} (8.3\%--12.6\% and 5.8\%--9.0\%, respectively), lower in \emph{D2} (5.0\%--12.6\% and 2.7\%--7.1\%), and lowest in \emph{D3} (0.3\%--4.7\% and 0.3\%--2.3\%). 
WebRTC fingerprinting is less prevalent but geographically uneven. It appears up to 15× more often in the UAE and Algeria than in the other three countries (e.g., 4.6\%, 4.4\%, and 0.3\%--0.8\% of \emph{D1} sites, respectively).

We identify 330 \emph{D1} sites (3.3\%) that execute WebRTC fingerprinting from Algeria or the UAE, but not from Germany, India, or the USA (e.g., \path{author.today} and \path{ria.ru}). 
In all these cases, third-party Yandex scripts invoke the WebRTC APIs (e.g., \path{mc.yandex.ru/metrika/tag.js} on 244 sites and \path{mc.yandex.ru/metrika/watch.js} on 40 sites).
These scripts also load from Germany, India, and the USA, but they do not execute WebRTC fingerprinting there. This suggests that the scripts condition WebRTC fingerprinting on the visitor’s location.

We manually inspect \path{mc.yandex.ru/metrika/tag.js} to identify this logic. 
The script implements WebRTC fingerprinting through an \path{RTCPeerConnection}.
A regional condition guards this routine: it executes only when the internal \path{isEU} state equals \path{0} (Listing~\ref{fig:yandex-webrtc}).
This condition likely reflects region-specific data-protection configurations, such as Yandex’s GDPR handling~\cite{yandex-gdpr}, and explains why the script does not fingerprint visitors from Germany. 
Unfortunately, we could not infer further its inactivity in India and the USA due to heavily obfuscated code

\begin{table}[t]
  \centering
  \scriptsize
  \caption{Fingerprinting prevalence (\% sites)}
  \label{tab:fp:all}
  \begin{tabular}{lrrrrr}
    \toprule
    & \textbf{Algeria} & \textbf{Germany} & \textbf{India} & \textbf{UAE} & \textbf{USA} \\
    \midrule
    \multicolumn{6}{l}{\textit{(a) Overall prevalence by website catalog}} \\
    \addlinespace[2pt]
    \quad \emph{D1} catalog & 15.4 & 14.0 &  9.9 & 16.7 & 13.5 \\
    \quad \emph{D2} catalog &  6.7 &  7.6 &  5.7 & 16.2 & 11.9 \\
    \quad \emph{D3} catalog &  0.3 &  3.3 &  1.4 &  5.1 &  4.7 \\
    \midrule
    \multicolumn{6}{l}{\textit{(b) Per-technique prevalence by website catalog}} \\
    \addlinespace[2pt]
    \multicolumn{6}{l}{\quad\textit{Audio}} \\
    \quad\quad \emph{D1} catalog & 8.1 & 9.0 & 5.8 & 8.4 & 8.7 \\
    \quad\quad \emph{D2} catalog & 6.7 & 4.1 & 2.7 & 6.5 & 7.1 \\
    \quad\quad \emph{D3} catalog & 0.3 & 2.2 & 0.8 & 2.0 & 2.3 \\
    \addlinespace[2pt] 
    \multicolumn{6}{l}{\quad\textit{Canvas image}} \\
    \quad\quad \emph{D1} catalog & 11.1 & 12.3 & 8.3 & 12.6 & 11.6 \\
    \quad\quad \emph{D2} catalog & 5.0 & 6.9 & 5.3 & 12.6 & 10.5 \\
    \quad\quad \emph{D3} catalog & 0.3 & 2.9 & 1.3 & 3.5 & 4.7 \\
    \addlinespace[2pt]
    \multicolumn{6}{l}{\quad\textit{WebRTC}} \\
    \quad\quad \emph{D1} catalog & 4.4 & 0.3 & 0.8 & 4.6 & 0.4 \\
    \quad\quad \emph{D2} catalog & 0.0 & 0.8 & 0.2 & 4.6 & 1.4 \\
    \quad\quad \emph{D3} catalog & 0.0 & 0.3 & 0.0 & 1.7 & 0.1 \\
    \addlinespace[2pt]

    \bottomrule
  \end{tabular}
\end{table}


\para{Statistical tests and validation.}
We use statistical analyses to validate the geographic differences observed in third-party tracking and fingerprinting. We examine whether differences in (1) tracking prevalence, (2) fingerprinting prevalence, and (3) the number of third-party trackers across locations are statistically significant. Table~\ref{tab:stat_tests} presents the results, with 95\% confidence intervals and effect sizes.

In \emph{D1}, we visit the same websites from all five vantage points. We therefore apply Cochran's Q test to tracking and fingerprinting prevalence and the Friedman test to per-site tracker counts; both account for repeated measurements on the same sites. All three differ significantly across vantage points: tracking prevalence ($Q(4)=82.0$, $p<.001$), fingerprinting prevalence ($Q(4)=426.5$, $p<.001$), and tracker counts (Friedman $Q(4)=1459.8$, $p<.001$). For tracking prevalence, the effect size is the small 3.1-point spread reported in Section~\ref{subsec:tracking}. This confirms that location barely changes \emph{whether} a globally popular site tracks. For fingerprinting prevalence, the 6.8-point spread from Section~\ref{subsec:fingerprinting} is a larger effect, which confirms that location does change whether fingerprinting occurs. For tracker counts, Kendall's $W=0.06$ indicates a small but systematic effect, in line with the per-site averages in Table~\ref{tab:tracking:summary}, where the UAE leads and India trails. The tests thus support that on the same websites, location has a small effect on how much tracking occurs and a negligible one on whether it occurs, and the large geographic differences lie in which trackers load.

In \emph{D3}, each vantage point contains a distinct set of websites. We therefore apply Pearson's chi-squared test to tracking and fingerprinting prevalence and the Kruskal--Wallis (KW) test to tracker-count distributions, both suited to independent samples. All three differ significantly across countries: tracking prevalence ($\chi^2(4)=1724.6$, $p<.001$), fingerprinting prevalence ($\chi^2(4)=285.3$, $p<.001$), and tracker counts ($H(4)=875.3$, $p<.001$). The confidence intervals show that the cross-country differences are not sampling noise: the intervals of the lowest and highest countries do not overlap for either tracking or fingerprinting. The effect sizes show that the differences are largest for tracking prevalence: Cramér's $V$ is $0.11$ for tracking, $0.05$ for fingerprinting, and $\varepsilon^2$ is $0.05$ for tracker counts.

\sa{\subsection{Server-side Lottery vs Geographic Influence}
\label{sec:lottery-tests}
Prior works showed that Web measurements may suffer from the \textit{lottery} effect due to differential server-side behavior~\cite{roth2022security}, and argued for basing findings on repeated observations. To separate geographic influence from this \textit{lottery} and quantify the variance, we ran the original (intra-) inconsistency test on the header data we collected across visits for D1 sites. This inconsistency remained low and consistent across vantage points (0.13\%--0.25\% of header observations), confirming that our crawling infrastructure itself is not a major source of variance (\eg~\textit{X-FO} ranges only 0.37\%--0.90\% across locations). 
When we instead compared the same headers across countries without separating out this server-side noise, the resulting variance was 5--10$\times$ larger (\eg~\textit{CSP} arose from 0.33\% to 5.2\%, \textit{HSTS} from 0.62\% to 3.6\%). However, once we isolated sites that returned identical headers on every repeated visit and measured only the remaining cross-country difference, the variance showed true geographic variance instead: \textit{CSP} still varied 2.35\%--3.37\% by country and \textit{HSTS} 1.48\%--1.73\%, revealing that a genuine, reproducible geographic effect persists on shipped headers even after eliminating server-side noises.

We extended the inconsistency tests to the inclusion and behavior of third-party scripts on D1 sites. Prior work studied the lottery effects for headers only, thus, no comparable baseline exists here. 
We noticed considerably higher inconsistencies and more variances across vantage points (15.13\%--22.85\% of sites showing a different inclusion set across visits), indicating that third-party script loading is inherently less deterministic than header deployment, independent of crawl location. 
Comparing script inclusion across countries without removing server-side jitters yields a 40.44\% variance overall (40.7\%--43.47\% per country), higher than for headers. 
After isolating these noises, the variance still remained sizeable: 19.99\%--26.18\% by country, ascertaining that geography meaningfully reshapes which third-party scripts a site loads. 
We further tested whether the \textit{behavior} of included script varies by geography, through analyzing JS API call categories (\eg~fingerprinting, network exfiltration, etc.). Cross-visit inconsistency was comparable for most categories (\eg~\textit{fingerprinting}: 0.12\%--0.49\% across vantage points), but higher for \textit{cookie storage} and \textit{eval} (7.23\%--20.62\%). Comparing the same signals across countries without noise separation showed substantially larger variance (\eg~\textit{cookie storage} rising to 40.9\% overall). After removing sites with noises, a true geographic artifact remained, with country-level \textit{cookie storage} variance ranging 37.53\%--48.33\% -- indicating that where users' cookies and their storage depends measurably geography, carrying direct privacy implications beyond mere script inclusions.}

\sa{Thus, any robust S\&P measurements on Web shall consider \textit{diverse} website catalog composition, as well as account for variances incurred due to (singular) location of observation to accurately scope and report their observations.}

\begin{listing}[t]
\caption{Simplified reconstruction of the regional condition controlling WebRTC fingerprinting in Yandex Metrika.}
\begin{lstlisting}[
      language=JavaScript,
      basicstyle=\ttfamily\scriptsize,
      frame=lines,
      breaklines=true,
      numbers=left,
      numberstyle=\tiny,
      xleftmargin=1.5em
  ]
function collectWebRTC(window, telemetry) {
  const PeerConnection = window.RTCPeerConnection 
  if (window.navigator.onLine && PeerConnection) {
    const pc = new PeerConnection({ iceServers: [] });
    // WebRTC fingerprinting
  }
}
if (state.get("isEU") === 0) { collectWebRTC(window, telemetry); } // Conditional call
\end{lstlisting}
\label{fig:yandex-webrtc}
\end{listing}


\section{Discussion \& Conclusion}
\label{sec:discussion}




\subsection{Limitations and Threats to Validity}
\label{sec:validity}

\para{Bot detection.}
Despite using realistic network environments and physical machines, sites may still detect automated crawlers and alter their behavior. Such cases may lead to under-reporting of the prevalent behaviors. While we resorted to best-effort strategy to filter any bot-induced noise during crawls and analysis, certain mechanisms (\eg~Cloudflare Challenges~\cite{cloudflare}) may have evaded our scrutiny.

\para{Instrumentation effects.}~
We used \pg{} to monitor the runtime characteristics of websites here. Naturally, our framework subsumes the inherent limitations of \pg{}. Specifically, we noticed page crashes and load timeouts on content-heavy pages, induced by \pg{} instrumentation. To mitigate this side effect, we visited each URL up to \numemp{5} times until successfully visiting them \numemp{3} times. In addition, we used data from \numemp{3} consistently successful crawls in our analysis to account for server-side randomness.  

\para{Consent-banner interaction.}~
Interacting with consent banners at scale is non-trivial. We used AutoConsent~\cite{AutoConsent}, which handles consent interfaces through site-specific rules and heuristics, following prior studies~\cite{moti2024targeted, zagi2025referrer, luo2025canvassing}. This remains best-effort. AutoConsent may miss or mishandle banners on some sites. We also attempt to accept all banners. Prior work~\cite{prasad2026regtrack} shows that acceptance markedly raises the number of trackers on EU-facing sites, while US-facing sites barely change. 
Our results thus reflect the \textit{post-consent} state.

\para{Incomplete detection and cloaking.}~
Finally, we note that filter lists, while an invaluable source for annotating tracking domains and continuously updated, are also often incomplete or inadequate for identifying all tracking domains in the wild, including regional trackers.
Thus, we believe that our overall findings only present a conservative lower bound of differences among websites across dimensions.

\para{Causal effects.}~Our comprehensive measurement design allowed us to identify trends that correlated with geographical variations. However, geography, as a whole, subsumes several other confounds (e.g., regional data regulations, development norms, digital market) that may be responsible for the observed trends and variations. While we discussed interesting cases and potential influencing factors for reported trends, unfortunately, identifying exact root cause(s) behind these variations would require us to test each of the potential sub-factors in isolation to arrive on concrete causal factors for individual trends. Notably, this is an inherent issue with measurement studies in general and warrants separate qualitative investigation. Hence, we refrained from presenting uneducated causal relations within the scope of our study and leave it for future work.

\para{Geographic and temporal limitations.} Through this study, we aimed to characterize the influence of geography on the user during browsing or measuring Web, through a representative set of websites and five distinct countries across the world. Naturally, our research design and technical requirements limited us to measure effects from five countries within a specific time period. Although these countries span multiple regions and account for a substantial share of global web traffic, we can neither realistically aim for nor claim to fully represent the diversity of the Internet here. Further, \textit{accurately} conducting cross–country measurements, as proposed in this work, introduces practical challenges. It requires physical machines in multiple countries. While we used academic infrastructure in some locations, we relied on personal machines connected to residential networks in others. This setup led to variability in network quality and computational resources, which affected crawl duration and scheduling. 
Nevertheless, we applied a best-effort strategy throughout and conducted all crawls within a relatively small time window, between March and May 2026. Despite these challenges, we find substantial evidence that isolated geography as an important influencing factor on the Web.

\subsection{Insights and Implications}

\para{Insights.} Our work highlights that users’ geographical location plays a significant role in shaping the S\&P risks they encounter on the web. We observe this effect in two distinct ways. First, the S\&P posture of regional websites intended for local users differs from that of globally popular websites. 
For third-party tracking, \emph{D1} and \emph{D2} sites exhibit similarly high prevalence, while \emph{D3} sites show lower and more variable prevalence across countries. 
For browser fingerprinting, the pattern is more monotonic: prevalence is highest in \emph{D1}, lower in \emph{D2}, and lowest in \emph{D3}. This suggests that measurements based only on globally popular websites may not reflect the tracking and fingerprinting behavior users encounter on regional and locally popular websites.

Second, we see geographic variation even when users access the \emph{same set of globally popular websites}. For identical sites in the \emph{D1} catalog, while tracking prevalence changes only modestly across countries, the number and set of third-party trackers loaded by the same websites often differ by vantage point. 
Similarly, fingerprinting varies across vantage points, with WebRTC fingerprinting showing a particularly clear geographic skew. 
In some cases we trace this variation to explicit region checks in first- and third-party code, which suggests that providers condition tracking on the visitor’s jurisdiction, in some cases to meet local privacy obligations, and serve the less protected version elsewhere.


Our investigation of geographic differences via script inclusion chain analysis also showed divergence, both at first- and third-party script levels. 
That is, differences in privacy-invasive behaviors, ad-tech market and prevalence, supply-chain or even security-critical inclusion pattern may arise at both first-party and third-party sources, when accessed through different vantage points. Such patterns may result from a range of factors, including location-dependent configuration, region-specific advertising practices, or the use of country/language-driven digital infrastructures. Lastly, we also report inconsistent security-related header deployment patterns by websites across regions and websites, offering differing security controls to users and potentially expose them to a myriad of security threats.
Consequently, the existing view of S\&P controls on the web, built largely from globally lenses and single vantage points, is partial.


\para{Implications.} These findings have important implications for multiple stakeholders within the web ecosystem. Web measurement researchers should move beyond globally-popular websites, single vantage point measurements, and VPN/cloud-based infrastructure, as these may mischaracterize the S\&P experiences of users who depend on local or regional services and provide a skewed view of the associated risks encountered worldwide. 
Filter lists maintainers and the ad-blocking community need to expand coverage by incorporating more regional domains and endpoints.
Security scanning tools that rely on known heuristics and behavioral signatures may also be affected by missing out capturing critical behavior, and thus need to perform evaluation under more realistic environments.
Web developers and service providers should audit region-dependent logic in their code and standardize security configurations to ensure that there are no unintentional variations across regions.
Finally, regulators and policymakers should account for geographic deviations in auditing and enforcement, as the same service may exhibit inconsistent S\&P controls across jurisdictions.

We argue that future web measurement studies and deployments must treat geographic diversity as a core dimension, despite the practical challenges involved in doing so. Without this, our understanding of web S\&P will remain incomplete. We hope that our findings motivate future work to broaden the geographic scope of web measurements.

\bibliographystyle{IEEEtran}
\bibliography{references}

\appendices
\section*{Ethics Considerations}
\label{sec:ethics}

We followed best practices for our Web measurements in line with the established standards of our community~\cite{kenneally2012menlo, kohno2023ethical}. We extensively evaluated the design, methodology, and the impact of our study before we began our data collection step. We detail our assessment in the following:

\textit{Crawling Websites \& Data Collection:}~We ensured little to no side effects to the websites we visited during our study as our crawlers only made, at most, ten GET requests during our entire study. Additionally, we also did not directly or indirectly interact with the visited website, beyond consenting to the cookie through injecting JavaScript, to limit any potential side effects. Thus, our crawls could not have led to any significant load or state-changing actions on the server, as we observe in practice. Further, we did not feed any fake or erroneous data to any of the websites, nor did we collect any personally-identifiable information at any stage. However, our crawls may violate the Terms of Service of websites that do not allow automated crawling. In this case, we resorted to performing cost-to-benefit analysis and arrived at the conclusion that the findings from our study provide a deeper understanding of the security and privacy postures of websites across geolocations; however, without raising any concerns for the websites or causing harm to individuals.

\textit{Crawling Infrastructure:} The machines, network, and other infrastructural components used for data collection and storage were either our personal devices or hosted within our university network infrastructure and entirely under our control. As such, we did not share the collected data or Web traffic with any third parties. For our preliminary experiments, we used \vpn{} to identify and understand the differential behavior of websites over VPN and actual headful machines. Here, we do not believe to have leaked any other information to the VPN beyond the list of websites we visited via the network. This does not raise any ethical concerns for the services or individuals, to our knowledge.

\section{Construction and Validation of the \emph{D2} Catalog}
\label{secc:app-d2-construction}

The \emph{D2} catalog contains country-coded versions of global websites. For example, \textit{example.dz} is a candidate for Algeria if a global version, such as \textit{example.com} or \textit{example.org}, appears in the global CrUX list. We build the catalog in two stages: candidate collection and validation.

We collect candidates by comparing each country's top 1 million CrUX sites with the global top 1 million. A site is a candidate if (1) it is not in \emph{D1}, (2) it uses one of the country's TLDs, and (3) its second-level domain appears in the global list under a different TLD. This yields 239 candidates for Algeria, 7,684 for Germany, 6,490 for India, 1,277 for the UAE, and 1,045 for the USA. We then validate them using two signals: favicon similarity and TLS certificates.

\para{Favicon Similarity.}
Favicons identify a website's brand. If a candidate and its global equivalent use visually similar favicons, they likely belong to the same brand. We collect favicons by parsing \texttt{<link rel="icon">} tags in the page HTML. We obtain favicons for 71\% of candidates in Algeria, 90\% in Germany, 75\% in India, 82\% in the UAE, 82\% in the USA, and for 83\% of their global equivalents. We normalize all favicons before comparison: we render them on a white background, pad them to squares while preserving aspect ratios, and resize them to $224 \times 224$ pixels. We encode each favicon with a CLIP ViT-B/32 image encoder and compute the cosine similarity between candidates and their global equivalents.

We validate a candidate if the similarity is at least 0.85. We tested several threshold values (0.70, 0.75, 0.80, 0.85, 0.90, and 0.95) and manually inspected 50 comparisons at each; 0.85 was the lowest threshold with no false positives in our sample. Favicon similarity validates 25\% of candidates in Algeria, 54\% in Germany, 23\% in India, 37\% in the UAE, and 35\% in the USA.

\para{TLS Certificates.}
TLS certificates provide another signal of shared ownership. Websites run by the same organization may share a certificate that covers several domains, or hold certificates issued to the same entity. We collect certificates for all candidates and their global equivalents and extract two fields: \textit{Subject Alternative Names (SANs)}, which list the domains a certificate covers, and \textit{organizationName}, which identifies the entity it was issued to. We retrieved certificates for 96\% of candidates in Algeria, Germany, and the UAE; 93\% in India; and 97\% in the USA.

We validate a candidate if (1) its global equivalent's hostname appears in the candidate's SANs, (2) the candidate's hostname appears in the equivalent's SANs, (3) both share an identical SAN set, or (4) both have a non-empty, matching \texttt{organizationName}. TLS validates 6.3\% of candidates in Algeria, 8.1\% in Germany, 2.3\% in India, 8.6\% in the UAE, and 1.8\% in the USA.

Combining both validations, the final \emph{D2} catalogs contain 65 sites for Algeria, 4,278 for Germany, 1,535 for India, 505 for the UAE, and 371 for the USA. Lower thresholds for favicon similarity or additional validation signals would enlarge them; we resorted to a high-confidence catalog of country-coded global websites for this study.

\section{Additional Tables \& Figures}
\label{app:add-tab-fig}

\clearpage
\onecolumn

\begin{table}[H]
\begin{minipage}[t]{0.48\textwidth}
    \centering
    \footnotesize
    \caption{Security-related HTTP Response Headers}
    \label{tab:sec-headers}
    \begin{tabular}{p{\dimexpr\linewidth-2\tabcolsep}}
    \toprule
    Access-Control-Allow-Credentials, Access-Control-Allow-Headers,
    Access-Control-Allow-Methods, Access-Control-Allow-Origin,
    Access-Control-Expose-Headers, Access-Control-Max-Age,
    Content-Security-Policy, Content-Security-Policy-Report-Only,
    Cross-Origin-Embedder-Policy, Cross-Origin-Opener-Policy,
    Cross-Origin-Resource-Policy, Expect-CT, Origin-Agent-Cluster,
    Permissions-Policy, Reporting-Endpoints, Report-To,
    Strict-Transport-Security, Timing-Allow-Origin,
    X-Content-Type-Options, X-Frame-Options,
    X-Permitted-Cross-Domain-Policies, X-XSS-Protection \\
    \bottomrule
    \end{tabular}
\end{minipage}\hfill
\begin{minipage}[t]{0.48\textwidth}
    \centering
    \footnotesize
    \caption{Country-specific eTLDs extracted from IANA and ICANN database for website cataloging.}
    \label{tab:etld-labels}
    \begin{tabular}{l p{\dimexpr\linewidth-2.2cm}}
    \toprule
    \textbf{Country} & \textbf{eTLDs} \\
    \midrule
    Algeria & \texttt{.dz} \\
    Germany & \texttt{.de}, \texttt{.bayern}, \texttt{.berlin}, \texttt{.hamburg}, \texttt{.nrw}, \texttt{.saarland}, \texttt{.cologne}, \texttt{.koeln} \\
    India   & \texttt{.in} \\
    UAE     & \texttt{.ae}, \texttt{.abudhabi}, \texttt{.dubai} \\
    USA     & \texttt{.us}, \texttt{.boston}, \texttt{.miami}, \texttt{.nyc} \\
    \bottomrule
    \end{tabular}
\end{minipage}
\end{table}

\begin{table}[H]
\begin{minipage}[t]{0.48\textwidth}
    \centering
    \scriptsize
    \setlength{\tabcolsep}{3pt}
    \caption{Top 10 third-party tracking domains across the combined D1/D2/D3 catalogs. Values are the percentage of sites on which each tracker was observed.}
    \label{tab:tracking:common:trackers}
    \begin{tabular}{lrrrrr}
    \toprule
    Tracker & Algeria & Germany & India & UAE & USA \\ \midrule
    \texttt{googletagmanager.com}   & 41.3 & 44.7 & 36.4 & 52.5 & 47.2 \\
    \texttt{google-analytics.com}   & 28.3 & 26.7 & 26.4 & 37.1 & 34.5 \\
    \texttt{doubleclick.net}        & 23.9 & 25.5 & 18.9 & 35.7 & 24.3 \\
    \texttt{google.com}             & 21.8 & 24.1 & 18.3 & 34.0 & 22.0 \\
    \texttt{googlesyndication.com}  & 15.4 & 19.3 & 11.0 & 11.9 & 11.9 \\
    \texttt{cloudflareinsights.com} & 14.4 &  6.0 & 12.2 & 14.4 & 16.2 \\
    \texttt{facebook.net}           & 10.2 & 13.6 &  8.9 & 17.6 & 10.0 \\
    \texttt{facebook.com}           &  9.0 & 12.7 &  7.6 & 17.0 &  9.1 \\
    \texttt{clarity.ms}             &  5.2 &  6.7 &  4.8 &  8.6 &  5.0 \\
    \texttt{googleapis.com}         &  5.7 &  3.8 &  3.7 & 11.9 &  5.0 \\
    \bottomrule
    \end{tabular}
\end{minipage}\hfill
\begin{minipage}[t]{0.48\textwidth}
    \centering
    \captionof{figure}{Prevalence (gain/loss count) of Top 10 third-party script sources on websites when accessed over VPN.}
    \label{fig:script-vpn}
    \includegraphics[width=\linewidth]{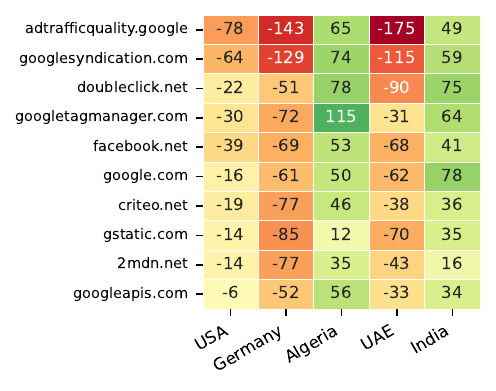}
\end{minipage}
\end{table}

\begin{table}[H]
    \centering
    \footnotesize
    \caption{Header Deployment over VPN: Comparison of deployment of Top 10 deployed security-related HTTP headers on Top \numemp{1500} globally popular websites (D1) when accessed with and without VPN. (Note that we could not collect VPN data from Algeria due to frequent network and electricity outages, and hence, we do not report them as of time of submission.)}
    \label{tab:vpn_headers}
    \begin{tabular}{l|c|c|c|c}
    \toprule
        HTTP Headers & GERMANY (VPN / Physical) & INDIA (VPN / Physical) & UAE (VPN / Physical) & USA (VPN / Physical) \\ \midrule
        Set-Cookie & 43.49\% / 46.70\% & 45.10\% / 44.83\% & 43.23\% / 45.99\% & 43.58\% / 47.77\% \\ \hline
        Strict-Transport-Security & 42.16\% / 44.56\% & 42.16\% / 40.11\% & 40.64\% / 42.60\% & 41.62\% / 44.74\% \\ \hline
        X-Content-Type-Options & 38.32\% / 40.11\% & 38.68\% / 37.43\% & 37.34\% / 40.29\% & 38.50\% / 40.82\% \\ \hline
        X-Frame-Options & 38.15\% / 39.66\% & 38.24\% / 36.27\% & 36.99\% / 38.41\% & 37.97\% / 39.84\% \\ \hline
        X-Xss-Protection & 28.61\% / 29.77\% & 29.14\% / 28.07\% & 28.61\% / 30.04\% & 28.88\% / 30.21\% \\ \hline
        Report-To & 26.65\% / 27.45\% & 26.92\% / 26.92\% & 25.76\% / 26.47\% & 26.65\% / 27.36\% \\ \hline
        Content-Security-Policy & 23.53\% / 25.04\% & 23.89\% / 23.35\% & 22.91\% / 24.60\% & 23.44\% / 25.40\% \\ \hline
        Access-Control-Allow-Origin & 9.63\% / 9.63\% & 9.00\% / 8.47\% & 9.27\% / 9.09\% & 9.54\% / 9.54\% \\ \hline
        Permissions-Policy & 6.77\% / 7.04\% & 6.86\% / 7.04\% & 6.60\% / 7.04\% & 7.31\% / 8.20\% \\ \hline
        Cross-Origin-Opener-Policy & 4.37\% / 4.63\% & 4.37\% / 4.19\% & 3.92\% / 5.17\% & 4.46\% / 4.63\% \\ \bottomrule
    \end{tabular}
\end{table}

\begin{table}[H]
    \centering
    \scriptsize
    \caption{Distribution of 3P-script \textit{request volume} loaded by websites across countries and catalogs \textit{(D1, D2, and D3)}.}
    \label{tab:3p_script_volume_distrib}
    \begin{tabular}{l|c|c|c|c|c|c|c|c}
    \toprule
        \textbf{Metric} & \textbf{URL Catalogs} & \textbf{ALGERIA} & \textbf{GERMANY} & \textbf{INDIA} & \textbf{UAE} & \textbf{USA} & \textbf{Mean} & \textbf{Std. Dev} \\ \midrule
        \multirow{3}{*}{Median 3P-scripts per site} & D1 & 105 & 97 & 42 & 132 & 39 & 83.00 & 40.92 \\
        ~ & D2 & 80.5 & 138 & 69 & 240 & 87 & 122.90 & 70.59 \\
        ~ & D3 & 45 & 66 & 27 & 115 & 43 & 59.20 & 34.14 \\
        \hline
        \multirow{3}{*}{Sites with $\geq$1 3P-scripts (\%)} & D1 & 92.54\% & 92.84\% & 92.46\% & 92.81\% & 92.93\% & 92.72\% & 0.20\% \\
        ~ & D2 & 86.79\% & 93.17\% & 90.54\% & 96.10\% & 79.55\% & 89.23\% & 6.40\% \\
        ~ & D3 & 66.24\% & 69.94\% & 67.43\% & 88.47\% & 86.33\% & 75.68\% & 10.81\% \\
        \hline
        \multirow{3}{*}{3P share of all script requests (\%)} & D1 & 73.61\% & 67.19\% & 59.59\% & 68.66\% & 63.28\% & 66.47\% & 5.33\% \\
        ~ & D2 & 71.07\% & 49.83\% & 67.65\% & 58.91\% & 59.43\% & 61.38\% & 8.31\% \\
        ~ & D3 & 42.87\% & 45.46\% & 48.14\% & 49.98\% & 55.01\% & 48.29\% & 4.62\% \\
    \bottomrule
    \end{tabular}
\end{table}

\begin{table}[H]
    \centering
    \scriptsize
    \caption{Effective policy (mis)configurations across HSTS, CSP, X-FO, and Set-Cookie header deployments (\emph{Security Lottery}), by URL category and country. Note: Since \pg{} only captures cookies visible at JS level, we do not report \textit{httpOnly}-related statistics here.}
    \label{tab:sec-lot-all}
    \begin{tabular}{l|l|c|c|c|c|c|c|c|c}
    \toprule
        Header & Flag & URL Cat. & ALGERIA & GERMANY & INDIA & UAE & USA & Mean & Std. Dev \\ \midrule
        \multirow{12}{*}{HSTS}
        & \multirow{3}{*}{No \textit{preload}} & D1 & 73.75\% & 73.11\% & 72.19\% & 73.10\% & 73.25\% & 73.08\% & 0.56\% \\
        & & D2 & 90.32\% & 75.41\% & 71.19\% & 71.96\% & 70.97\% & 75.97\% & 8.22\% \\
        & & D3 & 55.29\% & 81.09\% & 61.00\% & 73.09\% & 83.73\% & 70.84\% & 12.40\% \\ \cline{2-10}
        & \multirow{3}{*}{Missing \textit{includeSubDomains}} & D1 & 39.23\% & 37.88\% & 39.24\% & 38.59\% & 39.43\% & 38.87\% & 0.64\% \\
        & & D2 & 54.84\% & 43.00\% & 37.86\% & 42.06\% & 54.84\% & 46.52\% & 7.84\% \\
        & & D3 & 27.86\% & 41.58\% & 25.01\% & 35.78\% & 47.13\% & 35.47\% & 9.23\% \\ \cline{2-10}
        & \multirow{3}{*}{Non-standard \textit{max-age}} & D1 & 24.87\% & 24.71\% & 24.33\% & 24.86\% & 23.57\% & 24.47\% & 0.55\% \\
        & & D2 & 38.71\% & 22.75\% & 22.86\% & 24.77\% & 11.29\% & 24.08\% & 9.76\% \\
        & & D3 & 19.65\% & 19.63\% & 13.70\% & 21.69\% & 16.69\% & 18.27\% & 3.12\% \\ \cline{2-10}
        & \multirow{3}{*}{Disabled} & D1 & 1.60\% & 1.63\% & 1.82\% & 1.44\% & 1.62\% & 1.62\% & 0.13\% \\
        & & D2 & 0.00\% & 0.55\% & 0.00\% & 1.40\% & 0.00\% & 0.39\% & 0.61\% \\
        & & D3 & 0.22\% & 1.20\% & 2.19\% & 2.67\% & 3.54\% & 1.96\% & 1.29\% \\ \midrule
        \multirow{12}{*}{CSP}
        & \multirow{3}{*}{Prone to XSS} & D1 & 84.49\% & 85.92\% & 82.27\% & 86.02\% & 82.97\% & 84.33\% & 1.70\% \\
        & & D2 & 91.67\% & 92.65\% & 94.48\% & 94.40\% & 91.67\% & 92.97\% & 1.40\% \\
        & & D3 & 89.76\% & 88.01\% & 95.38\% & 97.08\% & 81.75\% & 90.40\% & 6.13\% \\ \cline{2-10}
        & \multirow{3}{*}{Prone to TLS attacks} & D1 & 71.21\% & 72.28\% & 73.11\% & 71.31\% & 75.60\% & 72.70\% & 1.80\% \\
        & & D2 & 75.00\% & 72.24\% & 46.51\% & 68.00\% & 33.33\% & 59.02\% & 18.21\% \\
        & & D3 & 76.10\% & 81.45\% & 65.94\% & 42.13\% & 68.88\% & 66.90\% & 15.12\% \\ \cline{2-10}
        & \multirow{3}{*}{Prone to framing attacks} & D1 & 39.81\% & 38.55\% & 40.30\% & 39.60\% & 37.75\% & 39.20\% & 1.03\% \\
        & & D2 & 50.00\% & 31.20\% & 58.43\% & 34.40\% & 22.22\% & 39.25\% & 14.68\% \\
        & & D3 & 50.24\% & 44.84\% & 56.10\% & 67.12\% & 35.03\% & 50.67\% & 12.03\% \\ \cline{2-10}
        & \multirow{3}{*}{Invalid policy} & D1 & 5.92\% & 5.85\% & 6.19\% & 5.68\% & 6.08\% & 5.94\% & 0.20\% \\
        & & D2 & 8.33\% & 11.51\% & 7.27\% & 10.40\% & 13.89\% & 10.28\% & 2.62\% \\
        & & D3 & 5.85\% & 7.35\% & 8.62\% & 6.03\% & 5.09\% & 6.59\% & 1.40\% \\ \midrule
        \multirow{9}{*}{X-FO}
        & \multirow{3}{*}{SELF} & D1 & 78.40\% & 78.14\% & 77.86\% & 78.92\% & 78.75\% & 78.41\% & 0.43\% \\
        & & D2 & 88.00\% & 79.84\% & 80.62\% & 86.10\% & 78.95\% & 82.70\% & 4.07\% \\
        & & D3 & 77.35\% & 73.77\% & 71.55\% & 82.48\% & 87.56\% & 78.54\% & 6.52\% \\ \cline{2-10}
        & \multirow{3}{*}{NONE} & D1 & 19.03\% & 19.36\% & 20.12\% & 18.48\% & 19.00\% & 19.19\% & 0.61\% \\
        & & D2 & 8.00\% & 16.57\% & 15.50\% & 11.23\% & 18.42\% & 13.94\% & 4.24\% \\
        & & D3 & 14.48\% & 23.22\% & 26.37\% & 16.01\% & 11.21\% & 18.26\% & 6.31\% \\ \cline{2-10}
        & \multirow{3}{*}{UNSAFE} & D1 & 2.57\% & 2.50\% & 2.02\% & 2.60\% & 2.25\% & 2.39\% & 0.25\% \\
        & & D2 & 4.00\% & 3.59\% & 3.88\% & 2.67\% & 2.63\% & 3.35\% & 0.66\% \\
        & & D3 & 8.17\% & 3.01\% & 2.07\% & 1.51\% & 1.23\% & 3.20\% & 2.86\% \\ \midrule
        \multirow{12}{*}{Set-Cookie}
        & \multirow{3}{*}{SameSite: Strict} & D1 & 0.16\% & 0.12\% & 0.19\% & 0.15\% & 0.20\% & 0.16\% & 0.03\% \\
        & & D2 & 0.00\% & 0.21\% & 0.63\% & 1.52\% & 0.00\% & 0.47\% & 0.64\% \\
        & & D3 & 0.08\% & 0.07\% & 0.10\% & 0.16\% & 0.15\% & 0.11\% & 0.04\% \\ \cline{2-10}
        & \multirow{3}{*}{SameSite: Lax} & D1 & 0.25\% & 0.24\% & 0.31\% & 0.28\% & 0.29\% & 0.27\% & 0.03\% \\
        & & D2 & 0.80\% & 0.81\% & 0.04\% & 0.14\% & 0.62\% & 0.48\% & 0.36\% \\
        & & D3 & 0.07\% & 0.87\% & 0.37\% & 0.09\% & 0.59\% & 0.40\% & 0.34\% \\ \cline{2-10}
        & \multirow{3}{*}{SameSite: None} & D1 & 99.59\% & 99.64\% & 99.50\% & 99.57\% & 99.51\% & 99.56\% & 0.06\% \\
        & & D2 & 99.20\% & 98.98\% & 99.33\% & 98.34\% & 99.38\% & 99.05\% & 0.43\% \\
        & & D3 & 99.85\% & 99.05\% & 99.52\% & 99.75\% & 99.27\% & 99.49\% & 0.33\% \\ \cline{2-10}
        & \multirow{3}{*}{\textit{secure} set to \textit{false}} & D1 & 99.69\% & 99.49\% & 99.14\% & 99.52\% & 99.39\% & 99.45\% & 0.20\% \\
        & & D2 & 99.99\% & 99.00\% & 99.58\% & 99.38\% & 99.22\% & 99.43\% & 0.38\% \\
        & & D3 & 99.83\% & 99.19\% & 99.53\% & 99.25\% & 99.74\% & 99.51\% & 0.29\% \\
        \bottomrule
    \end{tabular}
\end{table}

\begin{table}[H]
    \centering
    \scriptsize
    \caption{Statistical tests across vantage points for \emph{D1} (paired; Cochran's $Q$ for prevalence, Friedman for tracker counts, Kendall's $W$ as effect size) and \emph{D3} (independent; Pearson's $\chi^2$ with Cramér's $V$, Kruskal--Wallis with $\varepsilon^2$). Prevalence in \% with 95\% Wilson CIs; trackers per tracking site as median (IQR).}
    \label{tab:stat_tests}
    \setlength{\tabcolsep}{4pt}
    \begin{tabular}{lrrr}
    \toprule
    Country & 3P tracking (\%) & Fingerprinting (\%) & Trackers/site \\
    \midrule
    \multicolumn{4}{l}{\textit{(a) \emph{D1} catalog: same sites from every country}} \\
    Algeria & 71.4 [70.3, 72.5] & 15.4 [14.5, 16.3] & 5 (2--11) \\
    Germany & 70.2 [69.0, 71.3] & 14.0 [13.1, 14.8] & 6 (3--13) \\
    India   & 68.3 [67.1, 69.4] &  9.9 [9.2, 10.6]  & 4 (2--9) \\
    UAE     & 71.2 [70.1, 72.3] & 16.7 [15.8, 17.6] & 6 (3--15) \\
    USA     & 70.6 [69.5, 71.7] & 13.5 [12.6, 14.3] & 5 (2--12) \\
    \multicolumn{4}{l}{$Q(4)=82.0$ \quad $Q(4)=426.5$ \quad Friedman $Q(4)=1459.8$, $W=0.06$; all $p<.001$} \\
    \midrule
    \multicolumn{4}{l}{\textit{(b) \emph{D3} catalog: distinct sites per country}} \\
    Algeria & 35.0 [33.2, 36.8] & 0.3 [0.2, 0.6] & 2 (1--3) \\
    Germany & 50.8 [49.6, 51.9] & 3.3 [3.0, 3.8] & 3 (1--7) \\
    India   & 36.1 [35.1, 37.2] & 1.4 [1.2, 1.7] & 2 (2--5) \\
    UAE     & 65.0 [64.0, 66.0] & 5.1 [4.6, 5.5] & 4 (2--7) \\
    USA     & 55.4 [54.2, 56.6] & 4.7 [4.3, 5.3] & 3 (2--4) \\
    \multicolumn{4}{l}{$\chi^2(4)=1724.6$, $V=0.11$ \quad $\chi^2(4)=285.3$, $V=0.05$ \quad $H(4)=875.3$, $\varepsilon^2=0.05$; all $p<.001$} \\
    \bottomrule
    \end{tabular}
\end{table}

\end{document}